\documentclass[onecolumn,draftcls]{IEEEtran}

\usepackage{graphicx,pgfplots}
\usepackage{amsmath}
\usepackage{amsfonts,amssymb}
\usepackage{cite}
\usepackage{mathrsfs}
\usepackage{color}

\usepackage{tikz} 
\usetikzlibrary{arrows} 

\newtheorem{theorem}{Theorem}
\newtheorem{lemma}{Lemma}
\newtheorem{remark}{Remark}

\def\A{{\mathbf A}}

\def\C{{\mathbf C}}
\def\D{{\mathbf D}}

\def\R{{\mathbf R}}
\def\P{{\mathbf P}}

\def\I{{\mathbf I}}
\def\J{{\mathbf J}}
\def\L{{\mathbf L}}
\def\Q{{\mathbf Q}}
\def\F{{\mathbf F}}
\def\E{{\mathbf E}}
\def\U{{\mathbf U}}
\def\V{{\mathbf V}}
\def\G{{\mathbf G}}
\def\H{{\mathbf H}}
\def\S{{\mathbf S}}

\def\M{{\mathbf M}}

\def\e{{\mathbf e}}
\def\w{{\mathbf w}}

\def\x{{\mathbf x}}
\def\y{{\mathbf y}}
\def\z{{\mathbf z}}

\def\m{{\mathbf m}}

\def\d{{\mathbf d}}

\def\0{{\mathbf 0}}

\def\Gammab{{\boldsymbol \Gamma}}

\def\lambdab{{\boldsymbol \lambda}}
\def\Lambdab{{\boldsymbol \Lambda}}
\def\Psib{{\boldsymbol \Psi}}

\def\Omegab{{\boldsymbol \Omega}}

\def\Thetab{{\boldsymbol \Theta}}

\def\xib{{\boldsymbol \xi}}
\def\mub{{\boldsymbol \mu}}
\def\lambdab{{\boldsymbol \lambda}}

\def\Ht{\mathcal{H}_1}
\def\Hn{\mathcal{H}_0}

\def\pd{\mathbb{S}}
\def\dbplus{\mathbb{D}_{\mathrm{B}_+}}
\def\db{\mathbb{D}_{\mathrm{B}}}

\def\pmatrix{\mathbb{P}}

\DeclareMathOperator{\tr}{tr}

\DeclareMathOperator{\diag}{diag}

\providecommand{\re}{\mathrm{Re}}

\DeclareMathOperator{\nume}{e}

\newcommand{\detin}[1] {\det\left(#1\right)}
\newcommand{\jac}[1] {\left|\det\left(#1\right)\right|}
\newcommand{\trin}[1] {\tr\left(#1\right)}

\title{Locally Most Powerful Invariant Tests for Correlation and Sphericity of Gaussian Vectors}

\author{David~Ram\'irez,~\IEEEmembership{Member,~IEEE,}
        Javier~V\'ia,~\IEEEmembership{Member,~IEEE,} \\
        ~Ignacio~Santamar\'ia,~\IEEEmembership{Senior~Member,~IEEE}
        and~Louis~L.~Scharf,~\IEEEmembership{Life~Fellow,~IEEE}
\thanks{This work was presented in part at the 2012 IEEE International Conference on Acoustics, Speech, and Signal Processing (ICASSP) and at the 2012 IEEE Sensor Array and 
Multichannel Signal Processing Workshop (SAM).}
\thanks{The work of J. V\'ia and I. Santamar\'ia was supported by the Spanish Government, Mi\-nis\-terio de Ciencia e Innovaci{\'o}n (MICINN), under project COSIMA (TEC2010-19545-C04-03) and project COMONSENS (CSD2008-00010, CONSOLIDER-INGENIO 2010 Program). The work of L. Scharf was supported by the Airforce Office of Scientific Research under contract FA9550-10-1-0241.}
\thanks{D.~Ram\'irez is with the Signal and System Theory Group, Universit\"at Paderborn, 33098 Paderborn, Germany (e-mail: david.ramirez@sst.upb.de).}
\thanks{J.~V\'ia and I.~Santamar\'ia are with the Department of Communications
Engineering, University of Cantabria, 39005 Santander, Spain
(e-mail: \{jvia,nacho\}@gtas.dicom.unican.es).}
\thanks{L.~L.~Scharf is with the Departments of Mathematics and Statistics, Colorado State University, Ft. Collins, CO 80523,
USA (e-mail: Louis.Scharf@ColoState.edu).}
\thanks{Copyright (c) 2012 IEEE. Personal use of this material is permitted.  However, permission to use this material for any other purposes must be obtained from the IEEE by sending a request to pubs-permissions@ieee.org.}}

\begin{document}

\maketitle

\begin{abstract}

In this paper we study the existence of locally most powerful invariant tests (LMPIT) for the problem of testing the covariance structure of a set of Gaussian random vectors. The LMPIT is the optimal test for the case of close hypotheses, among those satisfying the invariances of the problem, and in practical scenarios can provide better performance than the typically used generalized likelihood ratio test (GLRT). The derivation of the LMPIT usually requires one to find the maximal invariant statistic for the detection problem and then derive its distribution under both hypotheses, which in general is a rather involved procedure. As an alternative, Wijsman's theorem provides the ratio of the maximal invariant densities without even finding an explicit expression for the maximal invariant. We first consider the problem of testing whether a set of $N$-dimensional Gaussian random vectors are uncorrelated or not, and show that the LMPIT is given by the Frobenius norm of the sample coherence matrix. Second, we study the case in which the vectors under the null hypothesis are uncorrelated \emph{and} identically distributed, that is, the sphericity test for Gaussian vectors, for which we show that the LMPIT is given by the Frobenius norm of a \emph{normalized} version of the sample covariance matrix.  Finally, some numerical examples illustrate the performance of the proposed tests, which provide better results than their GLRT counterparts.
\end{abstract}

\begin{keywords}
Hypothesis test, invariance, locally most powerful invariant test (LMPIT), maximal invariant statistic, Wijsman's theorem.
\end{keywords}

\section{Introduction}

Testing the covariance structure of Gaussian vectors is one of the classic problems in multivariate statistical analysis \cite{mardia_multivariate_analysis}, and is also commonly found in many applications, including sensor networks \cite{wireless_sensor_networks_information}, cooperative networks with multiple relays \cite{cooperation_wireless_network,laneman_cooperative,equalization_distributed_STBC_AF}, multiantenna radar detection \cite{mimo_radar} and, more recently, cognitive radio sensing with multiantenna spectrum monitors \cite{rankP_zeng_GC,rankP_zeng_TC,rank_P_trSP,roberto_rank_one_detection}. Assuming Gaussian vector measurements, the problem in its most general formulation consists of testing whether the covariance matrix of the stack of all vectors is block-diagonal, that is, whether all pairwise cross-covariance matrices are zero or not. A case of particular interest results when all vector measurements share the same covariance matrix under the null hypothesis, in which case the test is a test of sphericity for Gaussian vectors.

In practice, the likelihood depends on unknown parameters and the hypotheses are therefore composite. In the absence of a uniformly most powerful test (UMPT), most approaches consider the generalized likelihood ratio test (GLRT). However, despite its simplicity, it is known that the GLRT is not optimal in the Neyman-Pearson sense \cite{mardia_multivariate_analysis}. Furthermore, although the GLRT tends asymptotically to be optimal \cite{Kay_detection}, its performance may degrade for practical scenarios such as those typically found in cognitive radio applications, which are characterized by close hypotheses (low signal-to-noise ratio) and small sample sizes. When the problem exhibits symmetries or invariances, a reasonable approach consists in focusing on the class of tests satisfying the required invariances under a suitable group of transformations. Optimal invariant tests, which depend only on the maximal invariant statistic for the problem, are called uniformly most powerful invariant tests (UMPIT). Nevertheless, in many practical situations, such as those considered in this paper, the UMPIT does not exist (the maximal invariant statistic is vector-valued), and we have to resort to further refinements. Thus, by focusing on the challenging case of close hypotheses, and by applying a Taylor's series approximation of the ratio of maximal invariant densities, it might be possible to avoid the dependence on unknown parameters, yielding the so-called locally most powerful invariant test (LMPIT). The main goal of this paper is to study the existence of LMPITs for testing the covariance structure of Gaussian vectors, a problem for which UMP or UMPI tests do not exist in general.

 In contrast to the conventional way of deriving a LMPIT, which is finding the maximal invariant statistic and then deriving its distribution under both hypotheses, in this paper we apply Wijsman's theorem \cite{wijsman_theorem,eaton_book_1989,giri2004multivariate}. This powerful theorem (cf. Section \ref{sec:wijsman}) allows us to obtain the ratio of densities even without an explicit formulation for the maximal invariant statistic.

\subsection{Related works and main contributions}

The main theoretical work pertaining to the present paper was done in the seventies by John \cite{john_multivariate}, who derived the LMPIT in the scalar case for testing whether a set of real Gaussian random variables are uncorrelated and identically distributed (i.e., all with the same variance), the so-called sphericity test for Gaussian variables. Another related work is \cite{Schwartz_locally_minimax}, where the author derived a locally minimax test for testing independence among real Gaussian vectors. Apart from \cite{john_multivariate} and \cite{Schwartz_locally_minimax}, the vast majority of published work for testing the correlation structure of Gaussian random variables or random vectors with unknown parameters uses the GLRT. Specifically, the first generalized likelihood ratio test dates back to the work by Wilks in the 30s \cite{wilks_hadamard}, where the author derived the GLRT for testing 
the null hypothesis that the covariance matrix of a set of real Gaussian random variables is diagonal vs. the alternative that it is positive definite and otherwise arbitrary. That is a test for correlation among Gaussian variables. In particular, the GLRT derived by Wilks is given by the Hadamard ratio of the sample covariance matrix, i.e., the ratio between the determinant of the sample covariance matrix and the product of the elements of its main diagonal, which may also be rewritten as the determinant of the sample coherence matrix. More recently, this problem was revisited in the field of signal processing for radioastronomy by Leshem and Van der Veen \cite{leshem2001,vanderveen_detection_multichannel}, who derived the GLRT for circular complex Gaussian random variables. Interestingly, the determinant of the sample coherence matrix was also proposed in \cite{cochran_Trans_SP_95,GC_colored_noise} based on a geometric interpretation of the correlation coefficient, which they referred to as generalized coherence (GC). The GLRT for sphericity was found by Mauchly \cite{sphericity_mauchly}. These results have been extended to vector-valued Gaussian data in \cite{ramirez_GLRT_GCS_TrSP}, where the authors derived the GLRT for testing whether a covariance matrix of complex Gaussian vectors is block-diagonal or not. Not surprisingly, the GLRT for this problem is a \emph{generalized} (block-based) Hadamard ratio of the sample covariance matrix.

\begin{figure}[t]
\centering
\begin{tikzpicture}[scale=0.75,cap=butt]
  
  \draw[-triangle 60,black,thick] (-6,1.3) node[left,black] {$\x_1$}  -- (-3.7,1.3) ;
  \draw[-triangle 60,black,thick] (-6,-1.3) node[left,black] {$\x_L$} -- (-3.7,-1.3) ;
  
  \node at (-5.75,0) [circle,fill=black,rotate=90,align=center,inner sep=0pt,minimum size=1.5pt] {};
  \node at (-5.75,0.3) [circle,fill=black,rotate=90,align=center,inner sep=0pt,minimum size=1.5pt] {};
  \node at (-5.75,-0.3) [circle,fill=black,rotate=90,align=center,inner sep=0pt,minimum size=1.5pt] {};
    
  \node at (-3,1.375) [draw=none,fill=none,rotate=0,align=center] {$\hat{\R}_{11}$};
  \node at (-3,-1.235) [draw=none,fill=none,rotate=0,align=center] {$\hat{\R}_{LL}$};
  \draw[black,thick] (-3.7,-2) --  (-3.7,-0.6) -- (-2.3,-0.6) -- (-2.3,-2) -- cycle;
  \draw[black,thick] (-3.7,2) --  (-3.7,0.6) -- (-2.3,0.6) -- (-2.3,2) -- cycle;
  
  \draw[*-triangle 60,black,thick] (-5,-1.45) -- (-5,3.5) -- (3,3.5);
  \draw[*-triangle 60,black,thick] (-5.5,1.15) -- (-5.5,6.15) -- (3,6.15);
  
  \draw[*-triangle 60,black,thick] (-4.5,1.43) -- (-4.5,-3.5) -- (3,-3.5);
  \draw[*-triangle 60,black,thick] (-5.5,-1.17) -- (-5.5,-6.15) -- (3,-6.15);
  
  \draw[-triangle 60,black,thick] (-2.3,1.3) -- (-0.7,1.3) ;
  \draw[-triangle 60,black,thick] (-2.3,-1.3) -- (-0.7,-1.3) ;
  
  \node at (-1.1,0) [circle,fill=black,rotate=90,align=center,inner sep=0pt,minimum size=1.5pt] {};
  \node at (-1.1,0.3) [circle,fill=black,rotate=90,align=center,inner sep=0pt,minimum size=1.5pt] {};
  \node at (-1.1,-0.3) [circle,fill=black,rotate=90,align=center,inner sep=0pt,minimum size=1.5pt] {};
  
  \node at (0.15,0) [draw=none,fill=none,rotate=0,align=center] {$\displaystyle \frac{1}{L}\sum$};
  \draw[black,thick] (-0.7,-2) --  (-0.7,2) -- (1,2) -- (1,-2) -- cycle;
  
  \draw[black,thick] (3,6.85) --  (3,5.45) -- (4.55,5.45) -- (4.55,6.85) -- cycle;
  \draw[black,thick] (3,4.2) --  (3,2.8) -- (4.55,2.8) -- (4.55,4.2) -- cycle;
  \node at (3.78,6.15) [draw=none,fill=none,rotate=0,align=center] {Whiten};
  \node at (3.78,3.5) [draw=none,fill=none,rotate=0,align=center] {Whiten};
  
  \draw[-triangle 60,black,thick] (1,0) -- (1.9,0) -- (1.9,5.05) -- (3.7,5.05) -- (3.7,5.45) ;
  \draw[*-triangle 60,black,thick] (1.75,2.4) -- (3.7,2.4) -- (3.7,2.8) ;
  
  \draw[black,thick] (3,-6.85) --  (3,-5.45) -- (4.55,-5.45) -- (4.55,-6.85) -- cycle;
  \draw[black,thick] (3,-4.2) --  (3,-2.8) -- (4.55,-2.8) -- (4.55,-4.2) -- cycle;
  \node at (3.78,-6.15) [draw=none,fill=none,rotate=0,align=center] {Whiten};
  \node at (3.78,-3.5) [draw=none,fill=none,rotate=0,align=center] {Whiten};
  
  \draw[*-triangle 60,black,thick] (-2,-1.17) -- (-2,-5.05) -- (3.7,-5.05) -- (3.7,-5.45) ;
  \draw[*-triangle 60,black,thick] (-1.7,1.43) -- (-1.7,-2.4) -- (3.7,-2.4) -- (3.7,-2.8) ;
  
  \node at (7.2,-4.825) [draw=none,fill=none,rotate=0,align=center] {$\hat{\R}$};
  \draw[black,thick] (6.5,-6.85) --  (6.5,-2.8) -- (7.9,-2.8) -- (7.9,-6.85) -- cycle;
  
  \node at (5.8,-4.825) [circle,fill=black,rotate=90,align=center,inner sep=0pt,minimum size=1.5pt] {};
  \node at (5.8,-5.125) [circle,fill=black,rotate=90,align=center,inner sep=0pt,minimum size=1.5pt] {};
  \node at (5.8,-4.525) [circle,fill=black,rotate=90,align=center,inner sep=0pt,minimum size=1.5pt] {};
  
  \draw[-triangle 60,black,thick] (4.55,-6.15) -- (6.5,-6.15) ;
  \draw[-triangle 60,black,thick] (4.55,-3.5) -- (6.5,-3.5) ;
  
  \node at (7.2,4.825) [draw=none,fill=none,rotate=0,align=center] {$\hat{\R}$};
  \draw[black,thick] (6.5,6.85) --  (6.5,2.8) -- (7.9,2.8) -- (7.9,6.85) -- cycle;
  
  \node at (5.8,4.825) [circle,fill=black,rotate=90,align=center,inner sep=0pt,minimum size=1.5pt] {};
  \node at (5.8,5.125) [circle,fill=black,rotate=90,align=center,inner sep=0pt,minimum size=1.5pt] {};
  \node at (5.8,4.525) [circle,fill=black,rotate=90,align=center,inner sep=0pt,minimum size=1.5pt] {};
  
  \draw[-triangle 60,black,thick] (4.55,6.15) -- (6.5,6.15) ;
  \draw[-triangle 60,black,thick] (4.55,3.5) -- (6.5,3.5) ;
  
  \draw[-triangle 60,black,thick] (7.9,-4.825) -- (9.5,-4.825) ;
  
  \node at (10.3,-4.825) [draw=none,fill=none,rotate=0,align=center] {$\|\cdot\|^2$};
  \draw[black,thick] (9.5,-5.625) --  (9.5,-4.025) -- (11.1,-4.025) -- (11.1,-5.625) -- cycle;
  
  \draw[-triangle 60,black,thick] (11.1,-4.825) -- (12.4,-4.825) node[right,black] {LMPIT for Correlation} ;
  
  \draw[-triangle 60,black,thick] (7.9,4.825) -- (9.5,4.825) ;
  
  \node at (10.3,4.825) [draw=none,fill=none,rotate=0,align=center] {$\|\cdot\|^2$};
  \draw[black,thick] (9.5,5.625) --  (9.5,4.025) -- (11.1,4.025) -- (11.1,5.625) -- cycle;
  
  \draw[-triangle 60,black,thick] (11.1,4.825) -- (12.4,4.825) node[right,black] {LMPIT for Sphericity} ;
  
\end{tikzpicture}
\caption{Block diagram of the LMPIT for the testing problems considered in this paper. Whiten denotes a \emph{whitening step}, that is, the input vector is multiplied by the inverse of the square root matrix that enters into the block, $\hat{\R}$ is a block that computes the sample covariance matrix of the input-vectors, whereas $\hat{\R}_{kk}$ computes the covariance matrix of the $k$th vector, $1/L \sum$ averages the input matrices, and $\|\cdot\|^2$ denotes the squared Frobenius norm of the input-matrix. As can be seen, the main difference among the detectors is the whitening step. For testing correlation among Gaussian vectors, the whitening step is different for each vector, whereas for testing sphericity the same whitening is applied to each vector.}
\label{fig:block_diagram}
\end{figure}
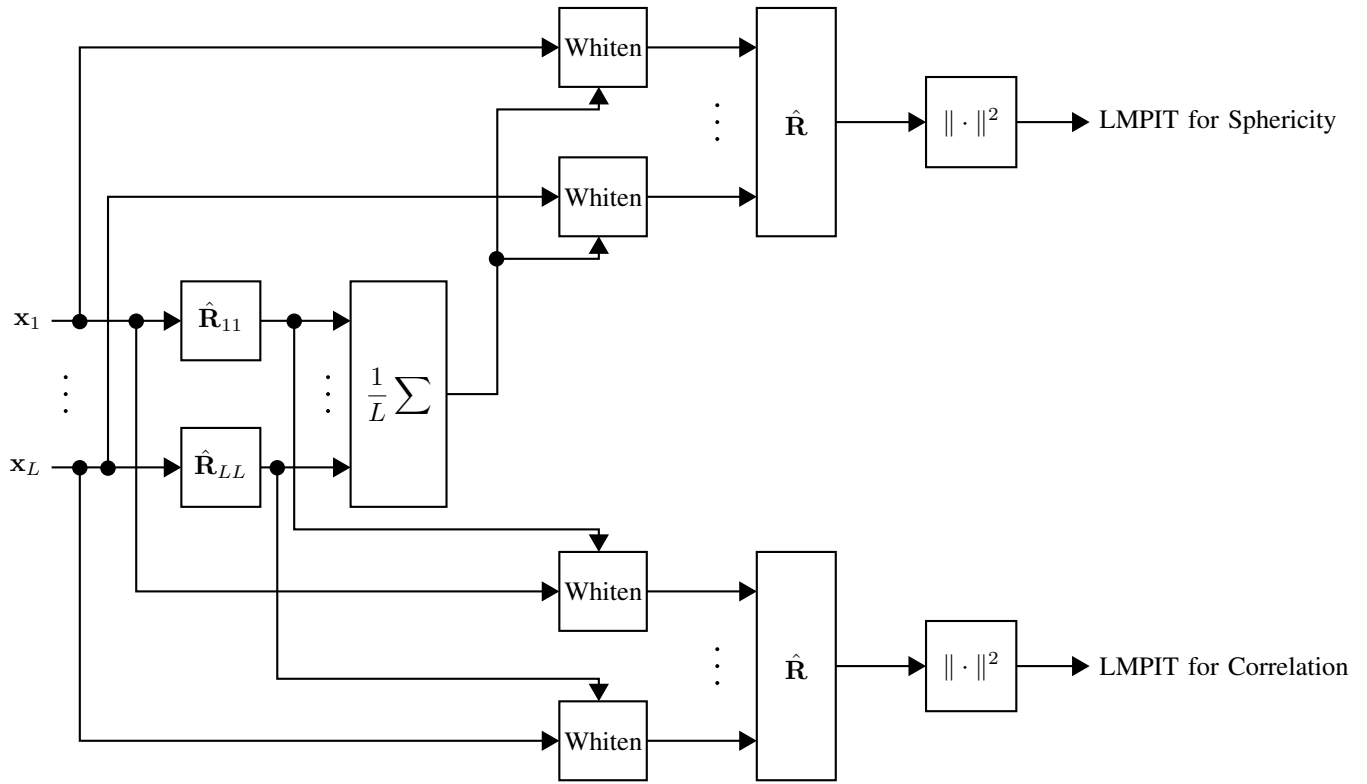

In this paper, we focus on the case of close hypotheses and study the existence of LMPITs for the covariance structure of Gaussian data. We extend \cite{john_multivariate} to the general case of vector-valued observations that may or may not have the same covariance matrix under the null hypothesis. Instrumental in deriving these results is the application of Wijsman's theorem, which requires identifying the invariances of each problem and integrating the distribution of the transformed observations under a measure on the corresponding group of transformations. Although we focus on the case of complex Gaussian vectors, which is motivated by its applications in radar and spectrum sensing problems, the presented results can be easily proved for real vectors, which is left as an exercise for the interested reader. Specifically, the main contributions of this work are the following:

\begin{itemize}
\item {\bf Correlation Test}: For vectors with different covariances under the null hypothesis, we show that the LMPIT is given by the Frobenius norm of the sample coherence matrix. The block diagram of the LMPIT is depicted in the lower part of Figure \ref{fig:block_diagram}, where we can see that the whitening is different for each vector. We should also point out that the Frobenius norm of the coherence matrix, or similar squared-sum test statistics, have been previously proposed as approximations of the GLRT that might be computationally simpler or have some other advantage \cite{leshem2001,vanderveen_detection_multichannel,correlation_lugosi,sample_EV_detection}. Here, we prove for the first time that the Frobenius norm of the coherence matrix is in fact the LMPIT for this problem.

\item {\bf Sphericity Test}: When the vector-valued observations under the null hypothesis are independent and identically distributed, we show that the LMPIT is given by the Frobenius norm of a \emph{normalized} version of the sample covariance matrix. Figure \ref{fig:block_diagram} also shows the block diagram of the LMPIT for this problem. However, in this case, the whitening is common for all vectors, and it is given by the inverse square root matrix of the average of the individual sample covariance matrices.

\item {\bf Local Irrelevance of the rank-structure}: Interestingly, the previous results also hold for the signal-plus-noise model commonly used in many signal processing problems, regardless of the rank of the signal covariance matrix. From a practical point of view, this means that for close hypotheses the spatial structure does not play any role for detection purposes. In other words, the use of subspace-based estimation techniques does not help for detection, at low signal-to-noise ratios (SNR) and/or low sample support.

\end{itemize}

The paper is organized as follows. Section \ref{sec:wijsman} presents a brief review of Wijsman's theorem and the use of invariance in detection problems. The LMPIT for testing whether $L$ non-identically distributed $N$-dimensional vectors are uncorrelated is introduced in Section \ref{sec:lmpit_noiid}, whereas Section \ref{sec:lmpit_iid} presents the LMPIT when the vector observations under the null hypothesis are identically distributed. The performance of the proposed detectors is illustrated by means of numerical simulations in Section \ref{sec:simulations}, and Section \ref{sec:conclusions} summarizes the main conclusions of this work.

\subsection{Notation}
\label{sec:notation}
In this paper we use bold-faced upper case letters to denote matrices, with elements $x_{k,l}$, bold-faced lower case letters for column vectors, and light-face lower case letters for scalar quantities. The superscripts $(\cdot)^T$ and $(\cdot)^H$ denote transpose and Hermitian transpose, respectively. The determinant, trace and Frobenius norm of a matrix ${\A}$ will be denoted, respectively, as $\detin{\A}$, $\trin{\A}$ and $\| \A \|$. The notation ${\A} \in \mathbb{C} ^{M \times N}$ $\left({\A} \in \mathbb{R} ^{M \times N}\right)$ will be used to denote a complex (real) matrix of dimension $M \times N$. For vectors, the notation ${\x} \in \mathbb{C} ^{M}$ $\left({\x} \in \mathbb{R} ^{M}\right)$ denotes a complex (real) vector of dimension $M$. The absolute value of the complex number $x$ is denoted as $|x|$ and $\x\sim \mathcal{CN}(\mub,\R)$ indicates that $\x$ is a complex circular Gaussian random vector of mean $\mub$ and covariance matrix $\R$. The expectation operator will be denoted as $E[\cdot]$ and $\otimes$ is the Kronecker product. $\I_L$ is the identity matrix of size $L \times L$ and $\0_L$ denotes the zero vector or the zero matrix (depending on the context) of sizes $L \times 1$ and $L \times L$, respectively. We use $\A^{1/2}$ ($\A^{-1/2}$) to denote the Hermitian square root matrix of the Hermitian matrix $\A$ ($\A^{-1}$). Finally, $\propto$ stands for equality up to additive and multiplicative positive constant (not depending on data) terms and $\text{diag}_N(\A)$ is a block-diagonal matrix formed by the $N \times N$ matrix blocks on the diagonal of $\A$, whereas $\text{diag}(\A)$ denotes the vector composed by the elements of the main diagonal of $\A$.

\section{Invariant Tests: Wijsman's Theorem}
\label{sec:wijsman}

It is well known that the Neyman-Pearson detector, given by the likelihood ratio, is the optimal test for simple hypotheses \cite{scharf_book}, that is, for hypotheses whose likelihood is known. Nevertheless, when the likelihood depends on unknown parameters, the hypotheses are composite and the problem is far from trivial. In fact, an optimal detector for composite hypothesis, known as a uniformly most powerful test (UMPT) \cite{scharf_book,lehmann_detection}, only exists in a few fortunate simple cases.

When a UMPT does not exist, a typical approach consists in focusing on the class of detectors invariant to some transformations suitable for the problem. Let us first introduce the concepts of invariance and maximal invariant statistic that will be used in the rest of the section. A detector is said to be invariant to the group of transformations $\mathcal{G}$ if \cite{scharf_book,ferguson_mathematical_statistics}
\begin{equation}
T(g(\x)) = T(\x), \quad \forall g(\cdot) \in \mathcal{G},
\end{equation}
where $T(\cdot)$ is the test statistic and $\x$ are the measurements. A statistic $M(\x)$ is maximal invariant if \cite{scharf_book,ferguson_mathematical_statistics}
\begin{equation}
M(g(\x)) = M(\x),
\end{equation}
for all $g(\cdot) \in \mathcal{G}$ and
\begin{equation}
M(\x_1) = M(\x_2),
\end{equation}
implies $\x_2 = g(\x_1)$ for some $g(\cdot) \in \mathcal{G}$. Therefore, the maximal invariant statistic organizes the measurements into sets that provide a constant value of $M(\x)$, and these sets are known as \emph{orbits}. Taking this into account, every invariant test may be written as a function of the maximal invariant statistic,
\begin{equation}
T(\x) = T(M(\x)),
\end{equation}
and we can therefore restrict our attention to rules that are functions of $M(\x)$. Using this important concept, the conventional approach to obtaining a uniformly most powerful invariant test (UMPIT) is as follows \cite{scharf_book}:
\begin{enumerate}
\item Identify the problem invariances and identify the corresponding transformation group.
\item Find the maximal invariant statistic.
\item Derive the densities of the maximal invariant statistic under each hypothesis.
\item Obtain the detector based on the ratio of the densities of the maximal invariant statistic.
\item Establish that the detector statistic has a monotone likelihood ratio
\end{enumerate}

This approach can be, for many problems, very difficult due to the need to obtain the distribution of the maximal invariant statistic under each hypothesis, which might itself be a complicated function of the data. Wijsman's theorem \cite{wijsman_theorem,eaton_book_1989,giri2004multivariate} provides an alternative (simpler) way to derive the ratio of densities of the maximal invariant statistic. The theorem states that, under some mild assumptions (presented later), the ratio of the maximal invariant densities is given by
\begin{equation}
\mathscr{L} = \frac{\int_{\mathcal{G} }p\left(g(\x); \Ht\right) \jac{\J_g} dg}{\int_{\mathcal{G} }p\left(g(\x); \Hn\right) \jac{\J_g} dg},
\label{eq:Wijsman}
\end{equation}
where $p\left( \x ; \mathcal{H}_i\right)$ is the probability density function (pdf) of $\x$ under the hypothesis $\mathcal{H}_i$, $\mathcal{G}$ is the group of invariant transformations, $\J_g$ denotes the Jacobian of the transformation $g(\cdot) \in \mathcal{G} $ and $d g$ is an invariant group measure, which in our problems may be taken as the usual Lebesgue measure. Despite its usefulness, Wijsman's theorem has received little attention in the information theory and signal processing communities and just a few works have exploited this result to derive invariant tests for particular problems \cite{kay_wijsman,GLRT_vs_UMPI_Kay,minimax_CFAR_wijsman,IRWUB_wijsman,LMPIT_quaterions,ICASSP_2012_LMPIT,SAM_2012_LMPIT}.

Stein \cite{stein_multivariate_1956} was the first to propose the idea of integrating over the group of transformations that describe the problem invariances. Later, the conditions for the validity of the theorem were studied in \cite{GLRT_vs_UMPI_Kay,wijsman_theorem,eaton_book_1989,wijsman1990invariant,wijsman_conditions,wijsman_conditions_correction} by Wijsman and other authors, who formally proved the theorem. For us, it suffices to consider the simplest set of conditions in \cite{wijsman_theorem} and \cite{GLRT_vs_UMPI_Kay}. In particular, $\mathcal{G}$ has to be a Lie group, a finite group or a composition of both and the observations have to belong to a linear Cartan $\mathcal{G}$-space, that is, a nonempty open subset (denoted as $\mathcal{S}$) of Euclidean space such that, for every $\x \in \mathcal{S}$, there exists a neighborhood $\mathcal{V}$ for which the closure of $\{g(\cdot) \in \mathcal{G} : g(\mathcal{V}) \cap \mathcal{V} \neq \emptyset\}$ is compact.

Unfortunately, for the testing problems considered in this paper, the ratio of the maximal invariant densities (obtained through Wijsman's theorem) still depends on an unknown parameter vector, which means that the UMPIT does not exist in general. For this reason, we will focus our attention on the case of close hypotheses and will study the existence of locally most powerful invariant tests (LMPIT). The main idea behind the LMPIT consists in applying a Taylor's series approximation of the likelihood ratio. When the lowest order term depending on the data is a monotone function of a scalar statistic, the detector is \emph{locally} optimal. We will detail this procedure in the following sections for the tests considered in the paper.


\section{The LMPIT for Correlation of Gaussian Vectors}
\label{sec:lmpit_noiid}

Consider a set of $L$ zero-mean circular complex jointly Gaussian vectors, $\x_1, \ldots, \x_L$, of dimension $N$, i.e., $\x_i \in \mathbb{C}^{N}, \ i = 1, \ldots, L$. In this section, we address the problem of testing whether these vectors are correlated or not, without assuming any particular structure for the covariance matrices. That is, we want to check if $E\left[\x_k \x_l^H \right] = \0_L$, $ \forall k \neq l$. We shall proceed by constructing the $NL$ vector $\x = [\x^T_1, \ldots, \x^T_L]^T$, and defining its covariance matrix as
\begin{equation}
\R = E[\x \x^H] = \begin{bmatrix}
\R_{11} & \R_{12} & \cdots & \R_{1L} \\
\R^H_{12} & \R_{22} & \cdots & \R_{2L} \\
\vdots & \vdots & \ddots & \vdots \\
\R^H_{1L} & \R^H_{2L} & \cdots & \R_{LL} \\
\end{bmatrix},
\end{equation}
with $\R \in \mathbb{C}^{L N \times L N}$ and $\R_{kl} \in \mathbb{C}^{N \times N}$. Hence, the detection problem may be cast as the following test for the covariance structure of $\x$,
\begin{equation}
\begin{array}{ll}
\Ht: \R \in \pd, \\
\Hn: \R \in \dbplus,
\end{array}
\end{equation}
where $\pd$ is the set of positive definite matrices\footnote{The results may be extended to consider positive semi-definite matrices.} and $\dbplus$ is the set of block-diagonal matrices with positive definite blocks. As previously pointed out, we do not impose any particular covariance structure on the individual vectors, beyond uncorrelation among vectors under $\Hn$. Then, the test becomes
\begin{equation}
\label{eq:test}
\begin{array}{ll}
\Ht: \x \sim \mathcal{CN}(\0_{LN},\R_1), \\
\Hn: \x \sim \mathcal{CN}(\0_{LN},\D),
\end{array}
\end{equation}
where $\R_1 \in \pd$ and $\D \in \dbplus$ are two unknown covariance matrices. This test is rather general and encompasses, for instance, the problem of testing whether a set of $L$ univariate non-stationary Gaussian time series are uncorrelated \cite{ramirez_GLRT_GCS_TrSP}.

\subsection{Derivation of the LMPIT}

Given a set of $M$ vector measurements, $\x[1], \ldots, \x[M]$, in this subsection we apply Wijsman's theorem to derive the LMPIT for the detection problem \eqref{eq:test}. To this end, it is first necessary to find the problem invariances. Specifically, this hypothesis test is invariant under the group of transformations
\begin{equation}
\label{eq:group}
\mathcal{G} = \left\{g: \x \rightarrow g(\x) = \P \G \x, \P = \overline{\P} \otimes \I_N, \overline{\P} \in \pmatrix, \G \in \db\right\}
\end{equation}
where $\pmatrix$ is the set of permutation matrices of dimension $L$ and $\db$ is the set of block diagonal invertible matrices. That is, the test is invariant under permutations of the vectors, and to a (possibly different) linear transformation of each of them.

Let us also obtain an explicit expression for the maximal invariant under the transformations in \eqref{eq:group}. This will help to identify the difficulties of the conventional approach to deriving invariant tests for this problem, and to further appreciate the value of Wijsman's theorem. It is well known that the sufficient statistic for this problem is the sample covariance matrix
\begin{equation}
\hat{\R} = \frac{1}{M} \sum_{m=1}^{M} \x[m] \x^H[m].
\end{equation}
Now, taking into account the invariance of the problem to a multiplication by an invertible block-diagonal matrix, we may introduce the transformation $\y = \hat{\D}^{-1/2} \x$, whose covariance matrix is
\begin{equation}
\hat{\R}_{\y \y} = \hat{\C} = \hat{\D}^{-1/2} \hat{\R} \hat{\D}^{-1/2} = \begin{bmatrix}
\I_N & \hat{\C}_{12} & \cdots & \hat{\C}_{1L} \\
\hat{\C}^H_{12} & \I_N & \cdots & \hat{\C}_{2L} \\
\vdots & \vdots & \ddots & \vdots \\
\hat{\C}^H_{1L} & \hat{\C}^H_{2L} & \cdots & \I_N \\
\end{bmatrix},
\end{equation}
where $\hat{\D} = \text{diag}_N(\hat{\R})$ and $\hat{\C}$ is the coherence matrix. After this initial prewhitening stage, we can arbitrarily fix the order of the vectors, for instance, according to the determinants of the matrices $\hat{\C}_{k,l}$.\footnote{The determinants will be different with probability one.} 

\begin{figure}[t]
\centering
\includegraphics[width=0.5\columnwidth]{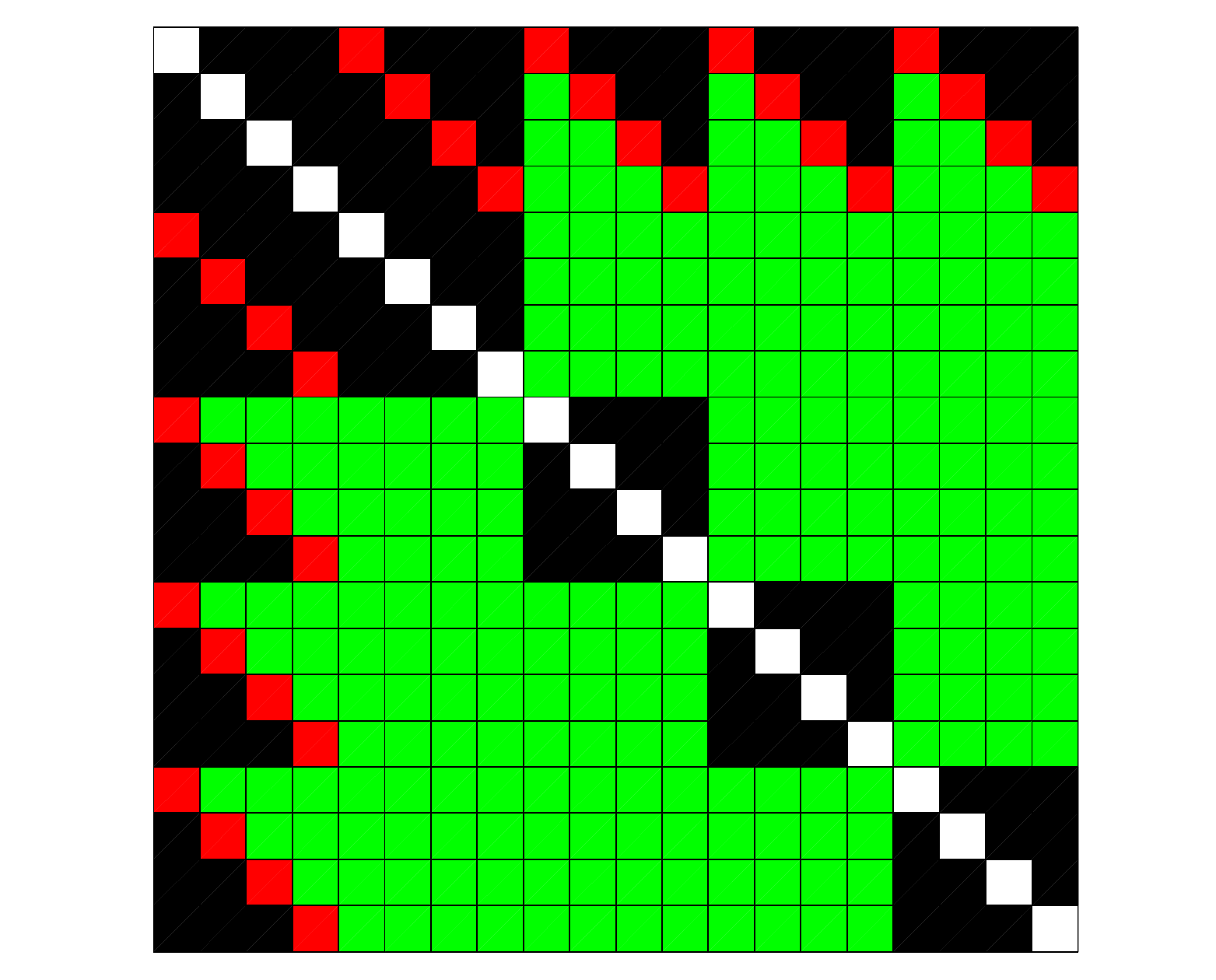}
\caption{Structure of $\R_{\w \w}$ (maximal invariant statistic) for the correlation test with $L=5$ vectors of size $N=4$. The black and white entries represent zeros and ones, respectively. Real positive and complex values are represented, respectively, by red and green entries.}
\label{fig:maximal_noniid}
\end{figure}

Once the order has been fixed, consider the singular value decomposition (SVD) of $\hat{\C}_{1,2}$
\begin{equation}
\hat{\C}_{12} = \U_{12} \text{diag}\left(\hat{\xib}_{12} \right) \V^H_{12},
\end{equation}
where $\hat{\xib}_{12}$ contains the canonical correlations between the first and second vectors. Thus, we may apply a unitary transformation of $\y$ belonging to $\mathcal{G}$, that diagonalizes $\hat{\C}_{12}$ as follows: $\z = \diag\left(\U_{12}^H,  \V_{12},  \I,  \ldots, \I \right) \y$. Additionally, the LQ decomposition of the first row and column partitions allows us to express $\hat{\R}_{\z \z}$ as
\begin{equation}
\hat{\R}_{\z \z} = \begin{bmatrix}
\I_N & \text{diag}(\hat{\xib}_{12}) & \L_{13} \Q_{13} & \cdots & \L_{1L} \Q_{1L} \\
\text{diag}(\hat{\xib}_{12}) & \I_N & \V_{12} \hat{\C}_{23} & \cdots & \V_{12} \hat{\C}_{2L} \\
\Q_{13}^H  \L_{13}^H & \hat{\C}_{23}^H \V_{12}^H & \I_N & \cdots & \hat{\C}_{3L} \\
\vdots & \vdots & \vdots & \ddots & \vdots \\
\Q_{1L}^H \L_{1L}^H & \hat{\C}_{2L}^H \V_{12}^H & \hat{\C}_{3L}^H & \cdots & \I_N \\
\end{bmatrix},
\end{equation}
where $\L_{kl}$ is a lower triangular matrix with non-negative real diagonal elements and $\Q_{kl}$ is a unitary matrix. Finally, we can apply another linear transformation to triangularize the first row and column partitions as follows $\w = \diag\left(\I, \I, \Q_{13}, \ldots, \Q_{1L} \right) \z$.
As an example, the final structure of $\hat{\R}_{\w \w}$ is shown in Figure \ref{fig:maximal_noniid}. The following lemma summarizes the obtained result.
\begin{lemma}
The maximal invariant statistic for the test \eqref{eq:test} under the group of transformations \eqref{eq:group} is given by
\begin{equation}
\left\{\hat{\xib}_{12}\right\} \cup \left\{\L_{1l}^H, \ l=3, \ldots, L\right\} \cup \left\{\Q_{1k} \hat{\C}_{kl} \Q_{1l}^H, \ k=2, \ldots, L, l = 3, \ldots, L, l > k\right\},
\end{equation}
where $\hat{\xib}_{12}$ are the canonical correlations between the first and second vectors, and $\L_{kl} \Q_{kl}$ is the LQ decomposition of $(k,l)$th block of $\hat{\R}_{\z \z}$.
\end{lemma}

Although we have been able to obtain an explicit expression for the maximal invariant statistic, it can be seen that it is a complicated function of the observations. Therefore, deriving its distribution under each hypothesis seems intractable. Additionally, Lemma 1 also allows us to conclude that, since for $N>1$ or $L>2$ the maximal invariant is a vector-valued function of the data, the UMPIT for this problem does not exist in general\cite{scharf_rank_one_detection}. Finally, notice that the maximal invariant statistic does not only depend on the eigenvalues of the coherence matrix (or canonical correlations), as shown in the following example.

\emph{Example:} Consider the following coherence matrices ($L =3, N = 1$)
\begin{equation*}
\hat{\C}_1 = \begin{bmatrix}
1 & 0.5 & 0 \\ 0.5 & 1 & 0 \\ 0 & 0 & 1
\end{bmatrix}, \quad
\hat{\C}_2 = \begin{bmatrix}
1& 0 & 0.4 \\ 0 & 1 & 0.3 \\ 0.4 & 0.3 & 1
\end{bmatrix},
\end{equation*}
which are not related by one of the transformations in the invariance group $\mathcal{G}$, i.e., they belong to different orbits. However, it can be easily verified that they share the eigenvalues (canonical correlations), given by $0.5, 1, 1.5$.
	
 Due to the difficulties posed by the maximal invariant of this problem, we resort to Wijsman's theorem to derive the LMPIT. It is easy to check that $\mathbb{D}_{\mathrm{B}}$ is a Lie group and, for $M \geq N$, the observation space $\mathcal{S} = \mathbb{C}^{L N \times M}$ is a linear Cartan $\mathcal{G}$-space. On the other hand, the group of permutations is finite and it is not therefore a Lie group. Then using the results in \cite{GLRT_vs_UMPI_Kay} for finite groups, and taking into account that also the conditions in \cite{wijsman_theorem} are fulfilled, Wijsman's theorem may be applied. Specifically, the ratio of the densities of the maximal invariant is given by
\begin{equation}
\label{eq:likelihood_ratio}
\mathscr{L} = \frac{\sum_{\pmatrix} \int_{\db } \detin{\R_1}^{-M} \jac{\G}^{2M} \nume^{ -M \trin{\P^T \R_1^{-1} \P \G \hat{\R} \G^H} } d \G }
{\sum_{\pmatrix} \int_{\db} \detin{\D}^{-M} \jac{\G}^{2 M}  \nume^{ -M \trin{\P^T \D^{-1} \P \G \hat{\R} \G^H} } d \G}.
\end{equation}
In order to derive the LMPIT, we must factor \eqref{eq:likelihood_ratio} as a function depending only on the observations and a function depending on the unknown parameters ($\R_1$ and $\D$). Before proceeding, let us introduce the following lemma.
\begin{lemma}
\label{lem:LR_NOIID}
The ratio of the maximal invariant densities \eqref{eq:likelihood_ratio} may be simplified as follows
\begin{equation}
\label{eq:likelihood_ratio3}
\mathscr{L} \propto \sum_{\pmatrix} \int_{\db } \jac{\G}^{2M} \nume^{ -M \trin{\G \G^H} } \nume^{ -\alpha } d \G,
\end{equation}
where
\begin{equation}
\alpha = M  \sum_{\substack{k,l = 1 \\ k \neq l}}^{L} \trin{\tilde{\S}_{kl} \G_{l} \hat{\C}_{lk} \G_{k}^H}.
\end{equation}
Here, the matrix $\G_{k}$ is the $k$th block of $\G$ and $\tilde{\S}_{kl}$ is the $(k,l)$th block of $\tilde{\S} = \P^T \D_{\S}^{-1/2} \S \D_{\S}^{-1/2} \P$, where $\S = \R_1^{-1}$ and $\D_{\S} = \text{diag}_N(\S)$.
\end{lemma}

\begin{IEEEproof}
 Considering the change of variable $\G \rightarrow \G \hat{\D}^{-1/2}$, which belongs to the group of transformations $\mathcal{G}$, \eqref{eq:likelihood_ratio} becomes
\begin{equation}
\label{eq:likelihood_ratio2}
\mathscr{L} \propto \frac{\sum_{\pmatrix} \int_{\db } \jac{\G}^{2M} \nume^{ -M \trin{\P^T \S \P \G \hat{\C} \G^H} } d \G }
{\sum_{\pmatrix} \int_{\db} \jac{\G}^{2 M}  \nume^{ -M \trin{\P^T \D^{-1} \P \G \hat{\C} \G^H} } d \G}.
\end{equation}
In words, this transformation shows that the detector does not depend on the individual sample covariances, but on the sample coherences.
Taking into account that the blocks of the main diagonal of $\hat{\C}$ are equal to the identity matrix, it is straightforward to prove that the denominator of \eqref{eq:likelihood_ratio2} does not depend on the observations. Finally, considering a new change of variable from the group of transformations $\G \rightarrow \P^T \D^{-1/2}_{\S} \P \G$, the result follows.
\end{IEEEproof}

It can be easily shown that, in general, \eqref{eq:likelihood_ratio3} is a function of the unknown parameters, given by $\S$, which allows us to conclude again that there does not exist a uniformly most powerful invariant test (UMPIT). The only exception is presented in the following theorem.
\begin{theorem}
For $L = 2$ and $N = 1$, that is, two scalar observations, the UMPIT\footnote{Actually, it is shown in \cite{lehmann_detection}, using a different approach, that the sample correlation coefficient is the uniformly most powerful unbiased test.} accepts $\Hn$ for small values of
\begin{equation}
T(\x[1], \ldots, \x[M]) = |\hat{c}_{12}|,
\end{equation}
where $\hat{c}_{12}$ is the sample correlation coefficient.
\end{theorem}
\begin{IEEEproof}
For scalar observations, \eqref{eq:likelihood_ratio3} is an increasing monotone function of the maximal invariant statistic $|\hat{c}_{12}|$.
\end{IEEEproof}
	
Due to the nonexistence of the UMPIT for vector-valued data, we shall focus on the challenging scenario of close hypotheses, and derive the locally most powerful invariant test (LMPIT), which is presented in the following theorem.
\begin{theorem}
\label{th:lmpit}
The LMPIT accepts $\Hn$ for small values of
\begin{equation}
\label{eq:lmpit}
T(\x[1], \ldots, \x[M]) =  \|\hat{\C}\|^2.
\end{equation}
\end{theorem}

\begin{IEEEproof}
See Appendix \ref{sec:proof_lmpit}.
\end{IEEEproof}

For the particular case of scalar observations, that is $N = 1$, the Frobenius norm of the coherence matrix has been previously proposed as an ad-hoc approximation of the GLRT detector \cite{leshem2001,vanderveen_detection_multichannel}. Additionally, in \cite{ramirez_GLRT_GCS_TrSP} the authors noticed experimentally that the Frobenius norm presents better performance than the determinant of the coherence matrix (which is the GLRT for this problem), mainly for close hypotheses and/or low sample sizes. Following a rigorous approach, in this section we have shown that the Frobenius norm of the coherence matrix is not just a reasonable approximation of the GLRT, but the LMPIT; that is, the optimal invariant detector for close hypotheses.

\subsection{Latent Signal Model}

Many signal processing, communications and econometrics problems use the well known signal-plus-noise or latent signal model. In these cases, the hypothesis test has further structure that, in principle, could be exploited to improve the detector performance. In particular, the test in this case is given by
\begin{equation}
\label{eq:test_latent}
\begin{array}{ll}
\Ht: \x \sim \mathcal{CN}(\0_{LN},\H \H^H + \D), \\
\Hn: \x \sim \mathcal{CN}(\0_{LN},\D),
\end{array}
\end{equation}
where $\H \in \mathbb{C}^{L N \times P}$ is the unknown channel matrix and $\D \in \mathbb{D}_{\mathrm{B}_+}$ is the unknown block-diagonal noise covariance matrix. The main difference with our original problem is that in \eqref{eq:test_latent} the covariance matrix under $\Ht$ has more structure than in \eqref{eq:test}. Specifically, it is the sum of a (possibly) rank-deficient matrix and the noise covariance matrix. As it turns out, the derivation of the GLRT for this problem is pretty involved, even for the simple case of $N = 1$ \cite{rank_P_trSP}. In particular, the authors of \cite{rank_P_trSP}, considering scalar observations, showed that there is no closed-form GLRT. On the other hand, it is easily proved that the additional structure imposed in $\R_1$ does not modify the invariances of the testing problem in \eqref{eq:test} and the maximal invariant remains the same, which has the following important consequence.
\begin{remark}
The LMPIT for the test given in \eqref{eq:test_latent} is also the Frobenius norm of the coherence matrix given by \eqref{eq:lmpit}, which means that, for close hypotheses, the additional spatial structure in the signal subspace is \emph{irrelevant} for optimal detection.
\end{remark}

The optimality of ignoring the spatial structure can be explained as a direct consequence of the fact that, for close hypotheses (e.g., low SNRs or small sample size), errors in the estimation of the signal subspace are likely to occur, thus degrading the performance of the detector. Consequently, it makes more sense in this situation to \emph{average out} the unknown parameters rather than using their maximum likelihood estimates.


\section{The LMPIT for Sphericity of Gaussian Vectors}
\label{sec:lmpit_iid}

In this section, we consider a more restrictive null hypothesis under which the vectors are also identically distributed. Specifically, we consider the following test
\begin{equation}
\label{eq:test_iid}
\begin{array}{ll}
\Ht: \x \sim \mathcal{CN}(\0_{LN},\R_1), \\
\Hn: \x \sim \mathcal{CN}(\0_{LN},\I_L \otimes \R_0),
\end{array}
\end{equation}
where $\R_1 \in \mathbb{S}$ is the $L N \times L N$ unknown covariance matrix under $\Ht$ and $\R_0 \in \mathbb{S}$ is the $N \times N$ unknown covariance matrix under $\Hn$. Hence, this hypothesis test generalizes the well-known test for sphericity of Gaussian variables \cite{sphericity_mauchly}. For this problem the group of invariant transformations is
\begin{equation}
\label{eq:group_iid}
\mathcal{G} = \left\{g: \x \rightarrow g(\x) = \left(\Q \otimes \G \right) \x, \Q \in \mathbb{Q}, \G \in \mathbb{G} \right\},
\end{equation}
where $\mathbb{Q}$ is the set of $L \times L$ unitary matrices and $\mathbb{G}$ is the set of $N \times N$ invertible matrices. That is, the problem is invariant under the \emph{same} linear transformation of all vectors and any unitary combination of them. The maximal invariant for this problem is again a complicated vector-valued function of the observations,\footnote{We leave the derivation of the maximal invariant statistic as an exercise for the interested reader.} which does not allow us to derive the distributions under each hypothesis and also shows that the UMPIT does not exist in general. In the following we use Wijsman's theorem to obtain the LMPIT for this problem.

The group of transformations $\mathcal{G}$ is a Lie group, and for $M \geq N$ the observation space $\mathcal{S} = \mathbb{C}^{L N \times M}$ is a linear Cartan $\mathcal{G}$-space. Hence, the conditions of Wijsman's theorem are fulfilled, which allows us to express the ratio of the maximal invariant densities as follows
\begin{equation}
\label{eq:likelihood_ratio_iid}
\mathscr{L} \propto \frac{\int_{\mathbb{Q}} \int_{\mathbb{G}}  \jac{\G}^{2LM} \nume^{ -M \trin{\R_1^{-1} \tilde{\G}  \hat{\R} \tilde{\G}^H} } d \G d\Q}
{\int_{\mathbb{Q}} \int_{\mathbb{G}} \jac{\G}^{2 L M}  \nume^{ -M \trin{ \left(\I \otimes \R_0^{-1} \right)  \tilde{\G} \hat{\R} \tilde{\G}^H} } d \G d\Q},
\end{equation}
where we have introduced the matrix $\tilde{\G} = \Q \otimes \G$ to simplify notation and dropped the size of the identity matrix. First, let us present a simpler expression for $\mathscr{L}$, given in the following lemma.
\begin{lemma}
The distribution ratio is given by
\begin{equation}
\label{eq:likelihood_ratio2_iid}
\mathscr{L} \propto \int_{\mathbb{Q}} \int_{\mathbb{G}} \jac{\G}^{2LM} \nume^{ -\alpha} d \G d\Q,
\end{equation}
where
\begin{equation}
\alpha = M \trin{\tilde{\S} \tilde{\G} \hat{\tilde{\R}} \tilde{\G}^H}.
\end{equation}
The \emph{normalized} sample covariance matrix is
\begin{equation}
\hat{\tilde{\R}} = (\I \otimes \hat{\R}_0)^{-1/2} \, \hat{\R} \,(\I \otimes \hat{\R}_0)^{-1/2},
\end{equation}
and the matrix $\tilde{\S}$ is given by
\begin{equation}
\tilde{\S} = (\I \otimes \overline{\S})^{-1/2} \, \S \,(\I \otimes \overline{\S})^{-1/2}, \quad \S = \R_1^{-1}.
\end{equation}
$\hat{\R}_0 = 1/L \sum_{k = 1}^{L} \hat{\R}_{kk}$ is the ML estimate of $\R_0$ and $\overline{\S} = 1/L \sum_{k = 1}^{L} \S_{kk}$ is the mean of the diagonal blocks of $\S$.
\end{lemma}

\begin{IEEEproof}
After applying the change of variable given by $\G \rightarrow \G \hat{\R}_0^{-1/2}$, the ratio \eqref{eq:likelihood_ratio_iid} becomes
\begin{equation}
\label{eq:likelihood_ratio3_iid}
\mathscr{L} \propto \frac{\int_{\mathbb{Q}} \int_{\mathbb{G}} \jac{\G}^{2LM} \nume^{ -M \trin{\S \tilde{\G} \hat{\tilde{\R}} \tilde{\G}^H} } d \G d\Q}
{\int_{\mathbb{Q}} \int_{\mathbb{G}} \jac{\G}^{2 L M}  \nume^{ -M \trin{ \left[ \I \otimes \left( \G^H \R_0^{-1} \G \right) \right]  \hat{\tilde{\R}}} } d \G d\Q}.
\end{equation}
To get rid of the denominator, let us apply in the integral of the denominator the change of variable $\G \rightarrow \R_0^{1/2} \G$, which yields
\begin{equation}
\label{eq:likelihood_ratio4_iid}
\mathscr{L} \propto \int_{\mathbb{Q}} \int_{\mathbb{G}} \jac{\G}^{2LM} \nume^{ -M \trin{\S \tilde{\G} \hat{\tilde{\R}} \tilde{\G}^H} } d \G d\Q.
\end{equation}
Finally, applying $\G \rightarrow \overline{\S}^{-1/2} \G$ concludes the proof.
\end{IEEEproof}

Analogously to the previous case, \eqref{eq:likelihood_ratio2_iid} is in general a function of the unknown parameters and, therefore, the UMPIT does not exist. In this case, the exception is presented in the following theorem, which was previously proved in \cite{scharf_rank_one_detection} following the conventional approach. 
\begin{theorem}
For $L = 2$ and $N = 1$, that is two scalar observations, the UMPIT accepts $\Hn$ for small values of
\begin{equation}
T(\x[1], \ldots, \x[M]) = \hat{\tilde{\xi}}_1,
\end{equation}
where $\hat{\tilde{\xi}}_1$ is the largest eigenvalue of $\hat{\tilde{\R}}$, or equivalently, the largest eigenvalue of $\hat{\R}$ normalized by $\tr(\hat{\R})$.
\end{theorem}
\begin{IEEEproof}
For $L = 2$ and $N = 1$, $\G$ becomes a scalar, namely $g$. Therefore, taking into account the eigenvalue decomposition (EVD) of $\tilde{\S}$ and $\hat{\tilde{\R}}$, the ratio \eqref{eq:likelihood_ratio2_iid} may be rewritten as
\begin{equation}
\label{eq:likelihood_ratio_iid_scalar}
\mathscr{L} \propto \int_{\mathbb{Q}} \int_{\mathbb{G}} |g|^{4M} \nume^{ -M |g|^2 \sum_{s,t}^{2} \tilde{\lambda}_t \hat{\tilde{\xi}}_s |q_{t,s}|^2 } d g d\Q,
\end{equation}
where $\tilde{\lambda}_t$ is the $t$th eigenvalue of $\tilde{\S}$ and $q_{t,s}$ denotes the $(t,s)$th element of $\Q$. Finally, noting that $\tr(\hat{\tilde{\R}}) = \hat{\tilde{\xi}}_1 + \hat{\tilde{\xi}}_2 = 2$, it is easy to show that \eqref{eq:likelihood_ratio_iid_scalar} is an increasing monotone function of $\hat{\tilde{\xi}}_1$.
\end{IEEEproof}

Excluding the particular case presented in the previous theorem, we have to focus on the case of close hypotheses and derive the LMPIT, which is presented next.
\begin{theorem}
\label{th:lmpit_iid}
The LMPIT accepts $\mathcal{H}_0$ for small values of 
\begin{equation}
\label{eq:lmpit_iid}
T(\x[1], \ldots, \x[M]) =  \|\hat{\tilde{\R}}\|^2.
\end{equation}
\end{theorem}
\begin{IEEEproof}
See Appendix \ref{sec:proof_lmpit_iid}.
\end{IEEEproof}

To conclude this subsection, let us consider again the latent signal model, which results in the following hypothesis test
\begin{equation}
\label{eq:test_iid_latent}
\begin{array}{ll}
\Ht: \x \sim \mathcal{CN}(\0_{L N},\H \H^H + \I_L \otimes \R_0 ), \\
\Hn: \x \sim \mathcal{CN}(\0_{L N}, \I_L \otimes \R_0).
\end{array}
\end{equation}
The group of invariant transformations for the above problem is still given by \eqref{eq:group_iid}. Hence, we can conclude that the LMPIT for the detection problem \eqref{eq:test_iid_latent} is also given by \eqref{eq:lmpit_iid}. This interesting result shows again that, under close hypotheses, the additional rank structure of the covariance matrix is \emph{irrelevant} for optimal detection.

\section{Numerical Results}
\label{sec:simulations}

In this section we illustrate the performance of the LMPITs in several scenarios. In all examples the covariance matrices are as follows
\begin{equation}
\begin{array}{ll}
\Ht: \R = \left(\F_L \Omegab_L \F_L^H\right) \otimes \I_{N}, \\
\Hn: \R = \I_{L N},
\end{array}
\end{equation}
where $\F_L$ is the Fourier matrix of dimensions $L \times L$ and $\Omegab_L$ is an $L \times L$ diagonal matrix containing the canonical correlations, i.e. the singular values of the coherence matrix. In particular, we have considered $L$ canonical correlations equispaced between $0.5$ and $1.5$. It is important to point out that, due to the problem invariances, there is no loss of generality by considering the identity covariance matrix under the null hypothesis.

\begin{figure}[t]
\centering
\input{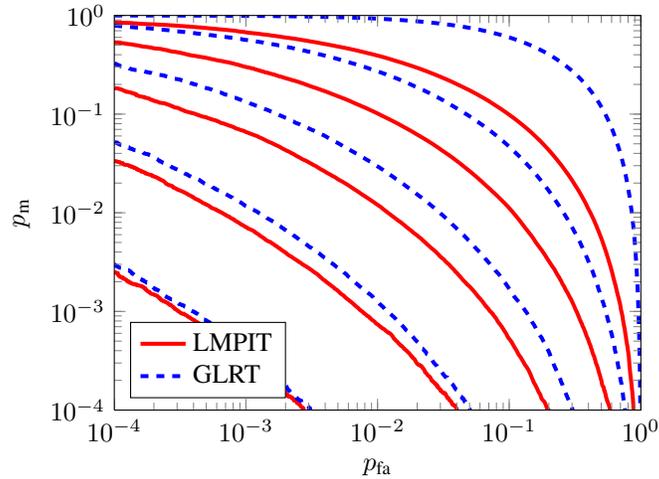}
\vspace*{-15pt}
\caption{ROC curves for the GLRT and the LMPIT for testing correlation among Gaussian vectors. The parameters of the experiment are: $L = 10$, $N = 4$ and $M = 40, 55, 70, 85,$ and $100$.}
\label{fig:noniid_vector}
\end{figure}

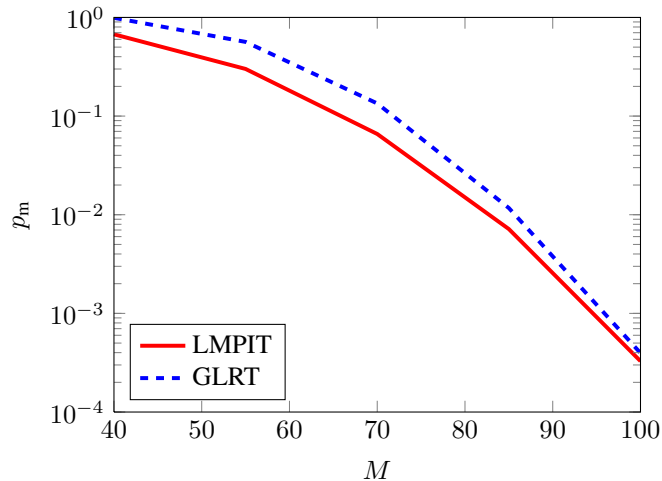
\begin{figure}[t]
\centering
%
%
\begin{tikzpicture}

\begin{semilogyaxis}[%
scale only axis,
width=7cm,
height=5.25cm,
xmin=40, xmax=100,
ymin=0.0001, ymax=1,
xlabel={$M$},
ylabel={$p_{\text{m}}$},
axis on top,
legend entries={LMPIT,GLRT},
legend style={at={(0.03,0.03)},anchor=south west,nodes=right}]

\addplot [
color=red,
solid,
line width=1.5pt
]
coordinates{
 (40,0.674498)(55,0.3004)(70,0.065842)(85,0.007158)(100,0.000328) 
};

\addplot [
color=blue,
dashed,
line width=1.5pt
]
coordinates{
 (40,0.990644)(55,0.564735)(70,0.134371)(85,0.011693)(100,0.000396) 
};

\end{semilogyaxis}
\end{tikzpicture}
\vspace*{-15pt}
\caption{Probability of missed detection vs. $M$ for the GLRT and the LMPIT for testing correlation among Gaussian vectors. The parameters of the experiment are: $L = 10$, $N = 4$ and we fix $p_{\text{fa}} = 10^{-3}$.}
\label{fig:pm_M_noniid_vector}
\end{figure}

In the first scenario we consider the performance of the LMPIT given by the Frobenius norm of the coherence matrix and compare it to that of the GLRT, which accepts $\Hn$ for large values of
\begin{equation}
T(\x[1], \ldots, \x[M]) = \log \detin{\hat{\C}}.
\end{equation}
Specifically, this example considers $L = 10$ vector measurements of dimension $N = 4$, and generates the  receiver operating characteristic (ROC) curves for $M = 40, 55, 70, 85,$ and $100$ samples. The ROC curves are shown in Figure \ref{fig:noniid_vector}, where we can see that the performance of the LMPIT is consistently better than that of the GLRT for this example with close hypotheses. Moreover, the performance gap increases for small values of $M$, as Figure \ref{fig:pm_M_noniid_vector} clearly shows.

\begin{figure}[t]
\centering
\input{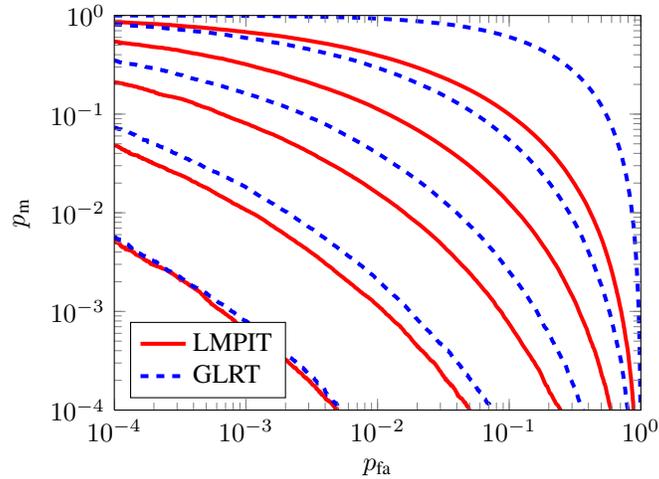}
\vspace*{-15pt}
\caption{ROC curves for the GLRT and the LMPIT for testing sphericity among Gaussian vectors. The parameters of the experiment are: $L = 10$, $N = 4$ and $M = 40, 55, 70, 85,$ and $100$.}
\label{fig:iid_vector}
\end{figure}

\begin{figure}[t]
\centering
%
%
\begin{tikzpicture}

\begin{semilogyaxis}[%
scale only axis,
width=7cm,
height=5.25cm,
xmin=40, xmax=100,
ymin=0.0001, ymax=1,
xlabel={$M$},
ylabel={$p_{\text{m}}$},
axis on top,
legend entries={LMPIT,GLRT},
legend style={at={(0.03,0.03)},anchor=south west,nodes=right}]
\addplot [
color=red,
solid,
line width=1.5pt
]
coordinates{
 (40,0.67869)(55,0.319582)(70,0.080544)(85,0.010699)(100,0.000649) 
};

\addplot [
color=blue,
dashed,
line width=1.5pt
]
coordinates{
 (40,0.990764)(55,0.59191)(70,0.161198)(85,0.018145)(100,0.000799) 
};

\end{semilogyaxis}
\end{tikzpicture}
\vspace*{-15pt}
\caption{Probability of missed detection vs. $M$ for the GLRT and the LMPIT for testing sphericity among Gaussian vectors. The parameters of the experiment are: $L = 10$, $N = 4$ and we fix $p_{\text{fa}} = 10^{-3}$.}
\label{fig:pm_M_iid_vector}
\end{figure}
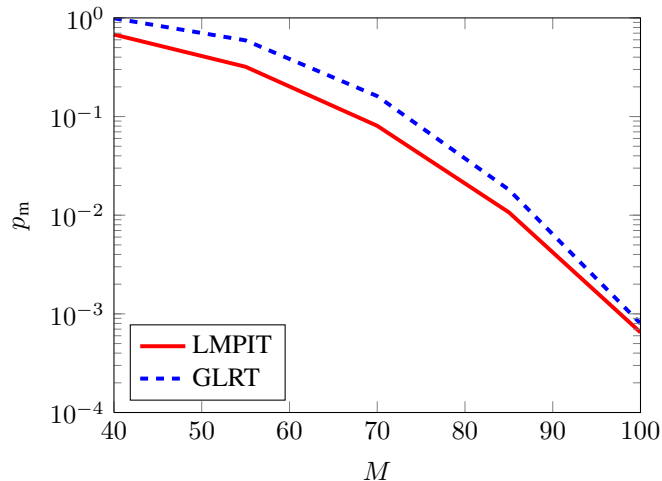

In the second scenario we evaluate the performance of the LMPIT given by \eqref{eq:lmpit_iid}, and that of the GLRT for the sphericity test for Gaussian vectors. In particular, the GLRT accepts $\Hn$ for large values of
\begin{equation}
T(\x[1], \ldots, \x[M]) =  \log \detin{\hat{\tilde{\R}}}.
\end{equation}
For this example, Figures \ref{fig:iid_vector} and \ref{fig:pm_M_iid_vector} show the ROC curves and the probability of missed detection, $p_{\text{m}}$, vs. $M$ for an experiment with $L = 10$ sets of vectors, $N = 4$ and $M = 40, 55, 70, 85,$ and $100$ observations, where we draw conclusions similar to those in the previous example.

\begin{figure}[t]
\centering
\input{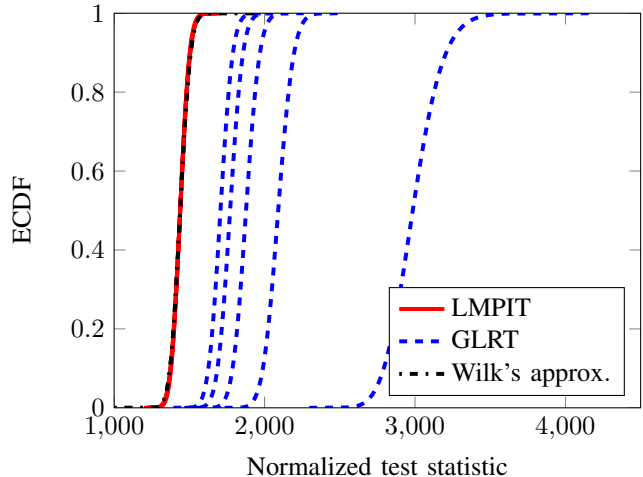}
\caption{Empirical CDF for the GLRT and the LMPIT for testing correlation among Gaussian vectors. The parameters of the experiment are: $L = 10$, $N = 4$ and $M = 40, 55, 70, 85,$ and $100$. Red solid line: LMPIT. Blue dashed line: GLRT (Curves far away from the Chi-squared approximation correspond to smaller values of $M$). Black dashed-dotted line: Wilks' approximation.}
\label{fig:hists_noniid_vector}
\end{figure}

Finally, let us make some comments on threshold selection. Similarly to \cite{Walden_propriety}, and exploiting the invariances of the problem, the thresholds may be obtained using numerical simulations for the particular case $\R_0 = \I$. These values will still be valid for any other choice of $\R_0$. As an example, Figure \ref{fig:hists_noniid_vector} shows the empirical cumulative distribution function of the LMPIT and the GLRT for the first scenario, where we can see that the distribution of the (normalized) LMPIT statistic is almost independent of $M$. Moreover, taking into account that $\log \det(\hat{\C}) \propto \| \hat{\C} \|^2$ for $\hat{\C} \approx \I$ \cite{leshem2001,ramirez_GLRT_GCS_TrSP}, we may use the results for the asymptotic distribution of the GLRT provided by Wilks' theorem \cite{Kay_detection}. Then, as $M \rightarrow \infty$, the distribution of the LMPIT converges to a Chi-squared distribution
\begin{align}
(M \|\hat{\C}\|^2 - L N M)  &\mathop{\sim}^{\Hn} \chi^2_{(L^2-L)N^2}, \\ (M \|\hat{\tilde{\R}}\|^2 - L N M) &\mathop{\sim}^{\Hn} \chi^2_{(L^2-1)N^2}.
\end{align}
In Figure \ref{fig:hists_noniid_vector} we can see how accurate the approximation is for this example.

\section{Conclusions}
\label{sec:conclusions}

In this paper, we have derived two locally optimal invariant tests for the covariance structure of Gaussian random vectors. We have focused on the case of close hypotheses, which is of interest in many practical problems such as cognitive radio sensing, where detection must be performed in low signal-to-noise ratio (SNR) scenarios. Locally most powerful invariant tests (LMPIT) for this problem were only known for the case of i.i.d. scalar observations (with the same variance under $\mathcal{H}_0$), since in this situation the maximal invariant statistic is a simple function of the observations and its distributions can be analytically characterized. For vector-valued data, the maximal invariant statistic is typically a complicated function of the observations. To avoid these difficulties, in this paper we have applied Wijsman's theorem to obtain the ratio of the maximal invariant densities by integrating over the group of transformations defining the invariances of the problem. Based on this theorem, we have proved that the LMPIT for testing whether a set of Gaussian vector measurements are correlated is given by the Frobenius norm of the sample coherence matrix. For random vectors identically distributed under the null hypothesis, that is, the sphericity test for Gaussian vectors, we have also proved that the LMPIT exists, and is given by the Frobenius norm of a \emph{normalized} sample covariance matrix. The invariances of the problems and our results do not depend on any additional rank structure the covariance matrices might have. In this way, an interesting conclusion from a practical point of view is that the additional spatial structure is locally irrelevant for optimal detection.

\appendices

\section{Proof of Theorem \ref{th:lmpit}}
\label{sec:proof_lmpit}

For close hypotheses (or low correlation), the inverse covariance matrix satisfies $\S \approx \I \Rightarrow \alpha \approx 0$. Therefore, applying a second order Taylor's series approximation of $\nume^{-\alpha}$ around $\alpha = 0$ we get
\begin{equation}
\label{eq:likelihood_ratio_approx1}
\mathscr{L} \propto \sum_{\pmatrix} \int_{\db } \beta(\G) \left(\alpha^2 - 2 \alpha\right)  d \G,
\end{equation}
where $\beta(\G) = \jac{\G}^{2M} \nume^{ -M \trin{\G \G^H} }$ is a positive term only depending on $\G$. Let us now consider the linear term, which is given by
\begin{equation}
\mathscr{L}_{\mathrm{l}} \propto \sum_{\pmatrix} \sum_{\substack{k,l = 1 \\ k \neq l}}^{L} \int_{\db } \beta(\G) \trin{\tilde{\S}_{kl} \G_{l} \hat{\C}_{lk} \G_{k}^H}  d \G.
\end{equation}
Applying the change of variable $\G_{k} \rightarrow - \G_{k}$, for all values of $k$, it is easy to check that all the integrals are equal to their opposites, so they must be zero. In consequence, the linear term must be zero. Using similar arguments, we can get rid of the cross-product terms in $\alpha^2$ and rewrite $\mathscr{L}$ as follows
\begin{equation}
\mathscr{L} \propto \sum_{\pmatrix}  \sum_{\substack{k,l = 1 \\ k \neq l}}^{L} \int_{\db } \beta(\G) \tr^2 \left(\tilde{\S}_{kl} \G_{l} \hat{\C}_{lk} \G_{k}^H \right)  d \G.
\label{eq:likelihood_ratio_approx2}
\end{equation}
 Performing the singular value decomposition (SVD) of $\tilde{\S}_{k,l}$ and $\hat{\C}_{l,k}$, the trace may be rewritten (without modifying the value of the integral) as
\begin{equation}
\trin{\tilde{\S}_{kl} \G_{l} \hat{\C}_{lk} \G_{k}^H} = \sum_{s, t = 1}^{N} \tilde{\lambda}^{kl}_t \hat{\xi}^{lk}_s g^{l}_{t,s} \left(g^{k}_{t,s}\right)^{\ast},
\end{equation}
where $\tilde{\lambda}^{kl}_t$ is the $t$th singular value of $\tilde{\S}_{kl}$ and $g^{l}_{t,s}$ denotes the $(t,s)$th element of $\G_l$. Therefore, \eqref{eq:likelihood_ratio_approx2} becomes
\begin{equation}
\mathscr{L} \propto \sum_{\pmatrix}  \sum_{\substack{k,l = 1 \\ k \neq l}}^{L} \int_{\db } \beta(\G) \left(\sum_{s, t = 1}^{N} \tilde{\lambda}^{kl}_t \hat{\xi}^{lk}_s g^{l}_{t,s} \left(g^{k}_{t,s}\right)^{\ast} \right)^2  d \G.
\end{equation}
Expanding the square, we get terms proportional to
\begin{equation}
g^{l}_{t,s} g^{l}_{t',s'} \left(g^{k}_{t,s} g^{k}_{t',s'}\right)^{\ast},
\end{equation}
and, using again a change of variable $g^{l}_{t,s} \rightarrow - g^{l}_{t,s}$ for $t \neq t'$ or $s \neq s'$, we can discard the cross terms, yielding
\begin{equation}
\mathscr{L} \propto \sum_{\pmatrix}  \sum_{\substack{k,l = 1 \\ k > l}}^{L} \sum_{s, t = 1}^{N} \left(\tilde{\lambda}^{kl}_t \hat{\xi}^{lk}_s \right)^2 \Delta,
\label{eq:likelihood_ratio_approx3}
\end{equation}
where
\begin{equation}
\Delta = \int_{\db } \beta(\G) \re \left[ \left(g^{l}_{t,s} \left(g^{k}_{t,s}\right)^{\ast} \right)^2 \right]  d \G.
\end{equation}
It is easy to see that $\Delta$ takes the same value for every required combination of the values of $k, l, s, t$ and, therefore, the ratio $\mathscr{L}$ becomes
\begin{equation}
\label{eq:likelihood_ratio_approx31}
\mathscr{L} \propto \sum_{\pmatrix}  \sum_{\substack{k,l = 1 \\ k > l}}^{L} \sum_{t = 1}^{N}  \left(\tilde{\lambda}^{kl}_t \right)^2 \sum_{s = 1}^{N}  \left( \hat{\xi}^{lk}_s \right)^2 = \sum_{\substack{k,l = 1 \\ k > l}}^{L}  \left[\sum_{\pmatrix} \left\| \tilde{\S}_{kl} \right\|^2 \right]  \left\|\hat{\C}_{lk}\right\|^2 .
\end{equation}
The sum over the set of permutations of the blocks of $\S$ cannot depend on the actual indexes, $k$ or $l$. In particular, recalling that the blocks of the main diagonal of $\tilde{\S}$ are equal to the identity, we have $\sum_{\pmatrix} \| \tilde{\S}_{kl} \|^2 \propto \| \tilde{\S} \|^2$, which allows us to write \eqref{eq:likelihood_ratio_approx31} as
\begin{equation}
\mathscr{L} \propto  \sum_{\substack{k,l = 1 \\ k > l}}^{L} \|\hat{\C}_{lk}\|^2 .
\end{equation}
Finally, we conclude the proof noting that $\|\hat{\C}_{lk}\|^2  = \|\hat{\C}_{kl}\|^2 $ and $\|\hat{\C}_{ll}\|^2  = N$.

\section{Proof of Theorem \ref{th:lmpit_iid}}
\label{sec:proof_lmpit_iid}

Focusing on the case of close hypotheses ($\tilde{\S} \simeq \I$) we can write
\begin{equation}
\mathscr{L} \propto \int_{\mathbb{Q}} \int_{\mathbb{G}} \beta(\G) (\alpha^2-2\alpha) \, d\G d\Q,
\end{equation}
where 
\begin{equation}
\alpha = M \trin{\tilde{\G}^H \overline{\S} \tilde{\G} \hat{\tilde{\R}} },
\end{equation}
and $\overline{\S} = \tilde{\S} - \I_{LN}$. Now, we shall consider the linear term, which may be written as
\begin{equation}
\int_{\mathbb{Q}} \int_{\mathbb{G}} \beta(\G) \trin{\tilde{\G}^H \overline{\S} \tilde{\G} \hat{\tilde{\R}} } \, d\G d\Q =  \trin{\Psib \hat{\tilde{\R}}},
\end{equation}
where
\begin{equation}
\Psib = \int_{\mathbb{Q}} \int_{\mathbb{G}} \beta(\G) \tilde{\G}^H \overline{\S} \tilde{\G} \, d\G d\Q.
\end{equation}

\begin{lemma}
	The matrix $\Psib$ is proportional to the identity.
\end{lemma}
\begin{IEEEproof}
	Consider the change of variable $\Q \rightarrow \Q \D$, where $\D = \text{diag}(\d)$ is one out of the $2^L$ different diagonal matrices with $\pm 1$ entries. Then, we can write
	\begin{equation}
	\Psib = \int_{\mathbb{Q}} \int_{\mathbb{G}} \beta(\G) \left(\D \otimes \I_N \right) \tilde{\G}^H  \overline{\S} \tilde{\G} \left(\D \otimes \I_N \right) d\G d\Q = \left(\D \otimes \I_N \right) \Psib \left(\D \otimes \I_N \right).
	\end{equation}
	Since the previous expression is valid for all the matrices $\D$, we can conclude that $\mathbf{\Psi}$ must be block-diagonal.
	Let us now consider the change of variable $\Q \rightarrow \Q \P$, where $\P \in \mathbb{P}$ is one out of the $L!$ different permutation matrices, yielding
	\begin{equation}
	\Psib = \int_{\mathbb{Q}} \int_{\mathbb{G}} \beta(\G) \left(\P \otimes \I_N \right) \tilde{\G}^H   \overline{\S} \tilde{\G} \left(\P \otimes \I_N \right)^T d\G d\Q = \left(\P \otimes \I_N \right) \Psib \left(\P \otimes \I_N \right)^T,
	\end{equation}
	and allows us to conclude that the blocks in the diagonal of $\mathbf{\Psi}$ are identical.
	In order to obtain a matrix proportional to the identity, we follow the previous lines exploiting the transformations $\G$. Therefore, introducing the change of variable $\G \rightarrow \G \D \P$, with $\D$ a diagonal matrix with $\pm 1$ entries and $\P$ a permutation matrix, we have
	\begin{equation}
	\Psib = \int_{\mathbb{Q}} \int_{\mathbb{G}} \beta(\G) \left(\I_L \otimes \D \P \right) \tilde{\G}^H   \overline{\S} \tilde{\G} \left(\I_L \otimes \D \P \right)^T d\G d\Q = \left(\I_L \otimes \D \P \right) \Psib \left(\I_L \otimes \D \P \right)^T,
	\end{equation}
	and we conclude that $\mathbf{\Psi} \propto \I_{LN}$.
\end{IEEEproof}

Now, taking into account the previous lemma and that $\tr(\hat{\tilde{\R}}) = L N$, it is easy to see that
\begin{equation}
\trin{\Psib \hat{\tilde{\R}}} \propto \trin{\hat{\tilde{\R}}} = L N,
\end{equation}
and, therefore, the linear term does not depend on data. Hence, the density ratio becomes
\begin{equation}
\mathscr{L} \propto \int_{\mathbb{Q}} \int_{\mathbb{G}} \beta(\G) \tr^2 \left(\tilde{\G}^H \overline{\S} \tilde{\G} \hat{\tilde{\R}} \right) \, d\G d\Q,
\end{equation}
which is simplified in the following lemma.

\begin{lemma}
\label{theorem:density}
	The density ratio of the maximal invariant statistic can be rewritten as
	\begin{align}
		\mathscr{L} &\propto	\int_{\mathbb{Q}} \int_{\mathbb{G}} \beta(\G) \tr^2 \left(\tilde{\G}^H \overline{\S} \tilde{\G} \hat{\tilde{\R}} \right) d\G d\Q  \nonumber \\ &=  \int_{\mathbb{Q}} \int_{\mathbb{G}} \beta(\G) \left[ \sum\limits_{k=1}^{L} \tr \left( \M_{kk} \hat{\tilde{\R}}_{kk} \right) \right]^2 d\G d\Q + 4 \sum\limits_{k=1}^{L} \sum\limits_{l=k+1}^{L} \int_{\mathbb{Q}} \int_{\mathbb{G}} \beta(\G) \re^2 \left\{ \tr \left( \M_{lk} \hat{\tilde{\R}}_{kl} \right) \right\} d\G d\Q, \label{eq:sumsquare}
	\end{align} 
	where $\M = \tilde{\G}^H \overline{\S} \tilde{\G}$.
\end{lemma}
\begin{IEEEproof}
	We shall decompose the trace as follows
	\begin{equation}
	\trin{ \tilde{\G}^H \tilde{\S} \tilde{\G} \hat{\tilde{\R}} } = \sum\limits_{k=1}^{L} \trin{\M_{kk} \hat{\tilde{\R}}_{kk} } + 2 \sum\limits_{k=1}^{L} \sum\limits_{l=k+1}^{L}\re \left\{ \trin{ \M_{lk} \hat{\tilde{\R}}_{kl} } \right\},
	\end{equation}
	and we may therefore write 
	\begin{equation}
		\int_{\mathbb{Q}} \int_{\mathbb{G}} \beta(\G) \tr^2 \left( \M \hat{\tilde{\R}} \right) d\G d\Q,
	\end{equation} 
	as a linear combination of terms of the form
	\begin{equation}
	\psi = \int_{\mathbb{Q}} \int_{\mathbb{G}} \beta(\G) \re \left\{ \trin{ \M_{mn} \hat{\tilde{\R}}_{nm} } \right\} \re \left\{ \trin{\M_{lk} \hat{\tilde{\R}}_{kl} } \right\} d\G d\Q.
	\end{equation}
	
	The above integrals can be simplified by introducing the change of variable $\Q \rightarrow \Q \D$ (again with $\D = \text{diag}(\d)$, and $\d = [d_1,\ldots,d_L]^T$ a vector with $\pm 1$ elements), which yields
	\begin{equation}
	\psi  = d_m d_n d_k d_l \int_{\mathbb{Q}} \int_{\mathbb{G}} \beta(\G) \re \left\{ \trin{ \M_{mn} \hat{\tilde{\R}}_{nm} } \right\} \re \left\{ \trin{\M_{lk} \hat{\tilde{\R}}_{kl} } \right\} d\G d\Q.
	\end{equation}
	Finally, the proof concludes by taking into account that the above equation must be valid for any choice of $\d$.
\end{IEEEproof}

Now, we shall split the proof in two parts, one for each term on the right hand side (RHS) of \eqref{eq:sumsquare}. The first is given in the following lemma.
\begin{lemma}
	The first integral in the RHS of \eqref{eq:sumsquare} is given by
	\begin{equation}
	\mathbf{\Gamma} \propto c \sum\limits_{k=1}^{L} \left\| \hat{\tilde{\R}}_{kk} \right\|^2,
	\end{equation}
	with $c \geq 0$ a constant term. \label{theorem:diag_blocks}
\end{lemma}
\begin{IEEEproof}
Considering the change of variable $\Q \rightarrow \Q \P_\mathbf{\pi}$, where $\P_\mathbf{\pi}$ is the permutation matrix defined by the re-ordering $\mathbf{\pi}$, we may write
	\begin{equation}
	\Gammab = \int_{\mathbb{Q}} \int_{\mathbb{G}} \beta(\G) \left[ \sum\limits_{k=1}^{L} \tr \left( \M_{kk} \hat{\tilde{\R}}_{kk} \right) \right]^2 d\G d\Q = \int_{\mathbb{Q}} \int_{\mathbb{G}} \beta(\G) \left[ \sum\limits_{k=1}^{L} \tr \left( \M_{kk} \hat{\tilde{\R}}_{\pi[k],\pi[k]} \right) \right]^2 d\G d\Q,
	\end{equation}
	where $\hat{\tilde{\R}}_{\pi[k],\pi[k]}$ denotes the $k$th block in the diagonal after permutation $\mathbf{\pi}$. Let us now expand the square
	\begin{multline}
	\Gammab = \int_{\mathbb{Q}} \int_{\mathbb{G}} \beta(\G)  \sum\limits_{k=1}^{L} \tr^2 \left( \M_{kk} \hat{\tilde{\R}}_{\pi[k],\pi[k]} \right) d\G d\Q  \\ + \int_{\mathbb{Q}} \int_{\mathbb{G}} \beta(\G)  \sum\limits_{k=1}^{L} \mathop{\sum\limits_{l=1}^{L}}_{l \neq k} \trin { \M_{kk} \hat{\tilde{\R}}_{\pi[k],\pi[k]} } \trin { \M_{ll} \hat{\tilde{\R}}_{\pi[l],\pi[l]} } d\G d\Q.
	\end{multline}
	Analogously to the previous cases, the value of the integral $\mathbf{\Gamma}$ does not depend on the particular choice of the permutation. Thus, we can average over the $L!$ possible permutations, and after some simplifications we get
	\begin{multline}
		\Gammab = \frac{1}{L} \int_{\mathbb{Q}} \int_{\mathbb{G}} \beta(\G)  \sum\limits_{k=1}^{L} \sum\limits_{m=1}^{L} \tr^2 \left( \M_{kk} \hat{\tilde{\R}}_{mm} \right) d\G d\Q \\
		+ \frac{1}{L^2-L} \int_{\mathbb{Q}} \int_{\mathbb{G}} \beta(\G)  \sum\limits_{k=1}^{L} \mathop{\sum\limits_{l=1}^{L}}_{l \neq k} \sum\limits_{m=1}^{L} \trin{ \M_{kk} \hat{\tilde{\R}}_{mm} } \trin{ \M_{ll} \left( L \I_N - \hat{\tilde{\R}}_{mm} \right) }  d\G d\Q,
	\end{multline}
	where we have used $\sum_{k=1}^{L} \hat{\tilde{\R}}_{kk} = L \I_N$. Expanding the second term yields
	\begin{multline}
		\Gammab = \frac{1}{L} \int_{\mathbb{Q}} \int_{\mathbb{G}} \beta(\G)  \sum\limits_{k=1}^{L} \sum\limits_{m=1}^{L} \tr^2 \left( \M_{kk} \hat{\tilde{\R}}_{mm} \right) d\G d\Q \\
		+ \frac{1}{L-1} \int_{\mathbb{Q}} \int_{\mathbb{G}} \beta(\G)  \sum\limits_{k=1}^{L} \mathop{\sum\limits_{l=1}^{L}}_{l \neq k} \sum\limits_{m=1}^{L} \trin{ \M_{kk} \hat{\tilde{\R}}_{mm} } \trin{ \M_{ll} }  d\G d\Q  \\
		 -\frac{1}{L^2-L} \int_{\mathbb{Q}} \int_{\mathbb{G}} \beta(\G)  \sum\limits_{k=1}^{L} \mathop{\sum\limits_{l=1}^{L}}_{l \neq k} \sum\limits_{m=1}^{L} \trin{ \M_{kk} \hat{\tilde{\R}}_{mm} } \trin{ \M_{ll} \hat{\tilde{\R}}_{mm} }  d\G d\Q,
	\end{multline}
	which, taking into account $\sum_{k=1}^{L} \hat{\R}_{kk} = L \I_N$ and defining $\overline{\M} = 1/L \sum_{k=1}^{L} \M_{kk}$, may be rewritten as
	\begin{multline}
		\Gammab \propto \frac{1}{L} \int_{\mathbb{Q}} \int_{\mathbb{G}} \beta(\G)  \sum\limits_{k=1}^{L} \sum\limits_{m=1}^{L} \tr^2 \left( \M_{kk} \hat{\tilde{\R}}_{mm} \right) d\G d\Q \\		
		- \frac{1}{L(L-1)} \int_{\mathbb{Q}} \int_{\mathbb{G}} \beta(\G)  \sum\limits_{k=1}^{L} \sum\limits_{m=1}^{L} \trin{ \M_{kk} \hat{\tilde{\R}}_{mm} } \trin{(L \overline{\M} -  \M_{kk}) \hat{\tilde{\R}}_{mm} }  d\G d\Q .
	\end{multline}
	Now, rearranging terms and summing in $k$, we get
	\begin{equation}
	         \label{eq:gamma_diff_traces}
		\Gammab \propto \frac{L}{L-1} \int_{\mathbb{Q}} \int_{\mathbb{G}} \beta(\G) \sum\limits_{m=1}^{L} \left[  \sum\limits_{k=1}^{L}  \tr^2 \left( \M_{kk} \hat{\tilde{\R}}_{mm} \right)  - \tr^2 \left( \overline{\M} \hat{\tilde{\R}}_{mm} \right)  \right] d\G d\Q.
	\end{equation}
	To conclude the proof of this lemma we must solve the above integral. To do so, let us consider the eigenvalue decomposition (EVD) of $\hat{\tilde{\R}}_{mm}$ as $\hat{\tilde{\R}}_{mm} = \V_m \Lambdab_m \V_m^H$ and introduce the change of variables $\G \rightarrow \G \V_m \P \D$ with $\P \in \mathbb{P}$ a permutation matrix, and $\D \in \mathbb{R}^{N \times N}$ a diagonal matrix with $\pm 1$ entries. Thus, we can write
	\begin{equation}
	\mathbf{\Gamma} \propto \frac{L}{L-1}  \int_{\mathbb{Q}} \int_{\mathbb{G}} \beta(\G)   \sum\limits_{m=1}^{L} \left[ \lambdab_{m}^T \P \D \left( \frac{1}{L} \sum\limits_{k=1}^{L} \m_k \m_k^T - \overline{\m} \hspace{30000sp} \overline{\m}^T \right) \D \P^T \lambdab_{m}   \right] d\G d\Q ,
	\end{equation}
	where $\overline{\m} = \text{diag}(\overline{\M})$, $\m_k = \text{diag}(\M_{kk})$, and $\lambdab_{m} = \text{diag}(\Lambdab_m)$ contains the eigenvalues of $\hat{\tilde{\R}}_{mm}$. Finally, $\mathbf{\Gamma}$ may be rewritten as
	\begin{equation}
	\mathbf{\Gamma} \propto   \sum\limits_{m=1}^{L} \lambdab_{m}^T  \E \lambdab_{m}, \label{eq:Gamma_quad}
	\end{equation}
	where
	\begin{equation}
	\E = \P \D \left[  \frac{L}{L-1}  \int_{\mathbb{Q}} \int_{\mathbb{G}}   \beta(\G) \left(\frac{1}{L} \sum\limits_{k=1}^{L} \e_k \e_k^T  \right)  d\G d\Q \right] \D \P^T,
	\end{equation}
	with $\e_k = \m_k - \m$, and we have taken into account that $\sum_{k=1}^L \m_k = L \m$. Thus, since the value of $\mathbf{\Gamma}$ does not depend on the particular choices of $\P$ or $\D$, we have
	\begin{equation}
	\E = \P \D \E \D \P^T = c \I_N,
	\end{equation}
	yielding
	\begin{equation}
	\mathbf{\Gamma} \propto c \sum\limits_{m=1}^{L} \left\| \hat{\tilde{\R}}_{mm} \right\|^2.
	\end{equation}
\end{IEEEproof}

The following lemma deals with the second integral in the RHS of \eqref{eq:sumsquare}.
\begin{lemma}
	The second integral in the expression of the density ratio can be rewritten as
	\begin{equation}
	4 \sum\limits_{k=1}^{L} \sum\limits_{l=k+1}^{L} \int_{\mathbb{Q}} \int_{\mathbb{G}} \beta(\G) \re^2 \left\{ \trin{ \M_{lk} \hat{\tilde{\R}}_{kl} } \right\} d\G d\Q = c \mathop{\sum\limits_{k,l=1}^{L}}_{l \neq k} \left\| \hat{\tilde{\R}}_{kl} \right\|^2,
	\end{equation}
	where $c$ is the same constant as in Lemma \ref{theorem:diag_blocks}. \label{theorem:offdiag_blocks}
\end{lemma}
\begin{IEEEproof}
Consider the integrals
\begin{equation}
	\Thetab_{k,l} = 4 \int_{\mathbb{Q}} \int_{\mathbb{G}} \beta(\G) \re^2 \left\{ \trin{ \M_{lk} \hat{\tilde{\R}}_{kl} } \right\} d\G d\Q,
\end{equation}
and the change of variable $\Q \rightarrow \Q \P$, with $\P$ a permutation matrix. Then, we can write
\begin{equation}
	\Thetab_{k,l} = 4 \int_{\mathbb{Q}} \int_{\mathbb{G}} \beta(\G) \re^2 \left\{ \trin{ \M_{\pi[l], \pi[k]} \hat{\tilde{\R}}_{kl} } \right\} d\G d\Q,
\end{equation}
and since the value of the integral cannot depend on the particular choice of the permutation, we may arbitrarily select 
\begin{equation}
	\Thetab_{k,l} = 4 \int_{\mathbb{Q}} \int_{\mathbb{G}} \beta(\G) \re^2 \left\{ \trin{ \M_{12} \hat{\tilde{\R}}_{kl} } \right\} d\G d\Q.
\end{equation}
Now, let us rewrite the previous integral as
\begin{equation}
	\Thetab_{k,l} = \int_{\mathbb{Q}} \int_{\mathbb{G}} \beta(\G) \tr^2 \left( \underline{\M} \hspace{30000sp} \underline{\R}_{kl} \right) d\G d\Q.
\end{equation}
where
\begin{equation}
	\underline{\M} = \begin{bmatrix}
	\M_{11} & \M_{12} \\
	\M_{12}^H & \M_{22}
	\end{bmatrix}, \qquad \underline{\R}_{kl} = \begin{bmatrix}
	\mathbf{0}_N & \hat{\tilde{\R}}_{kl}^H \\
	\hat{\tilde{\R}}_{kl} & \mathbf{0}_N
	\end{bmatrix}.
\end{equation}
We shall continue by applying the change of variable $\Q \leftarrow \Q \U$, where $\U$ combines the two first columns of $\Q$ as follows
\begin{equation}
	\U = \begin{bmatrix}
	\tilde{\U} & \0_{2 \times L-2} \\ \0_{L - 2 \times 2} & \I_{L-2}
	\end{bmatrix},
\end{equation}
with
\begin{equation}
	\tilde{\U} = \frac{1}{\sqrt{2}} \begin{bmatrix}
	1 & -1 \\
	1 & 1
	\end{bmatrix},
\end{equation}
and induces the transformation
\begin{equation}
	\underline{\R}_{kl} \leftarrow ( \tilde{\U} \otimes \I_N )^H \underline{\R}_{kl} ( \tilde{\U} \otimes \I_N ) = \frac{1}{2} \begin{bmatrix}
	\hat{\tilde{\R}}_{kl} + \hat{\tilde{\R}}_{kl}^H &  \hat{\tilde{\R}}_{kl}^H - \hat{\tilde{\R}}_{kl} \\
	\hat{\tilde{\R}}_{kl} - \hat{\tilde{\R}}_{kl}^H & - \hat{\tilde{\R}}_{kl} - \hat{\tilde{\R}}_{kl}^H
	\end{bmatrix}.
\end{equation}
The proof now follows the lines in Lemmas \ref{theorem:density} and \ref{theorem:diag_blocks}. First of all, as a direct consequence of Lemma \ref{theorem:density}, we may decompose $\Thetab_{k,l}$ as
\begin{equation}
	\Thetab_{k,l} = \frac{1}{4} \int_{\mathbb{Q}} \int_{\mathbb{G}} \beta(\G) \tr^2 \left( \left[\M_{11} - \M_{22} \right]  \left[ \hat{\tilde{\R}}_{kl} + \hat{\tilde{\R}}^H_{kl} \right]\right) d\G d\Q + \frac{1}{4} \int_{\mathbb{Q}} \int_{\mathbb{G}} \beta(\G) \tr^2 \left( \underline{\M} \overline{\R}_{kl} \right) d\G d\Q,
\end{equation}
with
\begin{equation}
         \overline{\R}_{kl} = \begin{bmatrix}
	\mathbf{0}_N & \hat{\tilde{\R}}_{kl}^H - \hat{\tilde{\R}}_{kl} \\
	\hat{\tilde{\R}}_{kl} - \hat{\tilde{\R}}_{kl}^H & \mathbf{0}_N
	\end{bmatrix}.
\end{equation}
Moreover, the second integral in the RHS of the previous equation can be simplified by introducing the change of variable $\Q \leftarrow \Q \U'$, with $\U'$ combining the two first columns by means of
\begin{equation}
	\tilde{\U}' = \frac{1}{\sqrt{2}} \begin{bmatrix}
	j & -j \\
	1 & 1
	\end{bmatrix}.
\end{equation}
This change of variable induces the transformation
\begin{equation}
	\overline{\R}_{kl} \leftarrow ( \tilde{\U}' \otimes \I_N )^H \overline{\R}_{kl} ( \tilde{\U}' \otimes \I_N ) = \begin{bmatrix}
	j (\hat{\tilde{\R}}_{kl} - \hat{\tilde{\R}}_{kl}^H) & \0_N \\
	\0_N & -j(\hat{\tilde{\R}}_{kl} - \hat{\tilde{\R}}_{kl}^H)
	\end{bmatrix},
\end{equation}
which yields
\begin{multline}
	\Thetab_{k,l} = \frac{1}{4}  \int_{\mathbb{Q}} \int_{\mathbb{G}} \beta(\G) \tr^2 \left( \left[\M_{11} - \M_{22} \right]  \left[ \hat{\tilde{\R}}_{kl} + \hat{\tilde{\R}}^H_{kl} \right]\right) d\G d\Q \\
	+ \frac{1}{4}  \int_{\mathbb{Q}} \int_{\mathbb{G}} \beta(\G) \tr^2 \left( \left[\M_{11} - \M_{22} \right] j \left[ \hat{\tilde{\R}}_{kl} - \hat{\tilde{\R}}^H_{kl} \right]\right) d\G d\Q.
\end{multline}
The rest of the proof follows the lines in Lemma \ref{theorem:diag_blocks}. The above integrals are almost\footnote{In this case, the sum of the diagonal blocks is zero, instead of $L \I_N$. However, this does not change the results.} identical in form to those considered in \eqref{eq:gamma_diff_traces} and therefore the same results apply. In particular, we have
\begin{equation}
	\Thetab_{k,l} =  \frac{c}{2} \left[  \left\| (\hat{\tilde{\R}}_{kl} + \hat{\tilde{\R}}_{kl}^H) \right\|^2 + \left\| j(\hat{\tilde{\R}}_{kl} - \hat{\tilde{\R}}_{kl}^H) \right\|^2 \right] = 2 c \left\| \hat{\tilde{\R}}_{kl} \right\|^2,
\end{equation}
	where $c \geq 0$ is the constant introduced in Lemma \ref{theorem:diag_blocks}. Finally, we can sum over all the off-diagonal blocks (instead of over those above the main diagonal) to obtain the desired result.
\end{IEEEproof}

Finally, the proof of the theorem follows from the direct combination of Lemmas \ref{theorem:density}, \ref{theorem:diag_blocks} and \ref{theorem:offdiag_blocks}.

\bibliographystyle{IEEEtran}
\bibliography{Mi_biblio}

\begin{IEEEbiographynophoto}{David Ram{\'\i}rez} (S'07-M'12)
received his Telecommunication Engineer Degree and his Ph.D. degree in electrical engineering from the University of Cantabria, Spain, in 2006 and 2011, respectively. From 2006 to 2011 he was with the Communications Engineering Department, University of Cantabria, Spain. In 2011, he joined the University of Paderborn, Germany, where he is currently a research associate. He has been a visiting researcher at the University of Newcastle, Australia. His current research interests include signal processing for wireless communications, MIMO Systems, and multivariate statistical analysis. Dr. Ram{\'\i}rez has been involved in several national and international research projects on these topics.
\end{IEEEbiographynophoto}

\begin{IEEEbiographynophoto}{Javier V{\'\i}a}
(S'04-M'08) received his Telecommunication Engineer Degree and his Ph.D. in
electrical engineering from the University of Cantabria, Spain, in
2002 and 2007, respectively. In 2002 he joined the Department of
Communications Engineering, University of Cantabria, Spain, where he
is currently Assistant Professor. He has spent visiting periods at
the Smart Antennas Research Group of Stanford University, and at the
Department of Electronics and Computer Engineering (Hong Kong
University of Science and Technology). Dr. V{\'\i}a has actively
participated in several European and Spanish research projects. His
current research interests include blind channel estimation and
equalization in wireless communication systems, multivariate
statistical analysis, quaternion signal processing and kernel
methods.
\end{IEEEbiographynophoto} 

\begin{IEEEbiographynophoto}{Ignacio Santamar{\'\i}a} (M'96-SM'05)
received his Telecommunication Engineer Degree and his Ph.D. in
electrical engineering from the Universidad Polit\'{e}cnica de
Madrid (UPM), Spain, in 1991 and 1995, respectively. In 1992 he
joined the Departamento de Ingenier{\'\i}a de Comunicaciones,
Universidad de Cantabria, Spain, where he is currently Full
Professor. He has been a visiting researcher at the Computational
NeuroEngineering Laboratory (University of Florida), and at the
Wireless Networking and Communications Group (University of Texas at
Austin). He has more than 100 publications in refereed journals and
international conference papers. His current research interests
include signal processing algorithms for wireless communication
systems, MIMO systems, multivariate statistical techniques and
machine learning theories. He has been involved in several national
and international research projects on these topics. He is currently
serving as a member of the Machine Learning for Signal Processing
Technical Committee of the IEEE Signal Processing Society.
\end{IEEEbiographynophoto}

\begin{IEEEbiographynophoto}{Louis L. Scharf} (S'67-M'69-SM'77-F'86-LF'07) received the Ph.D. degree from the University of Washington, Seattle.

From 1971 to 1982, he served as Professor of electrical engineering and statistics at Colorado State University (CSU), Ft. Collins. From 1982 to 1985, he was Professor and Chairman of electrical and computer engineering at the University of Rhode Island, Kingston. From 1985 to 2000, he was Professor of electrical and computer engineering at the University of Colorado, Boulder. In January
2001, he rejoined CSU as Professor of electrical and computer engineering and statistics. He has held several visiting positions here and abroad, including the Ecole Superieure d'electricit{\'e}, Gif-sur-Yvette, France; Ecole Nationale Superieure des T{\'e}l{\'e}communications, Paris, France; EURECOM, Nice, France; the University of La Plata, La Plata, Argentina; Duke University, Durham, NC; the University of Wisconsin, Madison; and the University of Troms{\o}, Troms{\o}, Norway. His interests are in statistical signal processing, as it applies to adaptive radar, sonar, and wireless communication. His most important contributed to date are to invariance theories for detection and estimation; matched and adaptive subspace detectors and estimators for radar, sonar, and data communication; and canonical decompositions for reduced dimensional filtering and quantizing. His current interests are in rapidly adaptive receiver design for space-time and frequency-time signal processing in the radar/sonar and wireless communication channels.

Prof. Scharf was Technical Program Chair for the 1980 IEEE International Conference on Acoustics, Speech, and Signal Processing (ICASSP), Denver, CO; Tutorials Chair for ICASSP 2001, Salt Lake City, UT; and Technical Program Chair for the Asilomar Conference on Signals, Systems, and Computers 2002. He is past-Chair of the Fellow Committee for the IEEE Signal Processing Society and serves on it Technical Committee for Sensor Arrays and Multichannel Signal Processing. He has received numerous awards for his research contributions to statistical signal processing, including a College Research Award, an IEEE Distinguished Lectureship, an IEEE Third Millennium Medal, and the Technical Achievement and Society Awards from the IEEE Signal Processing Society.
\end{IEEEbiographynophoto}

\end{document}